\documentclass[prd,aps,showpacs,floatfix,superscriptaddress,notitlepage,onecolumn]{revtex4-1}
\usepackage{graphicx}
\usepackage[ansinew]{inputenc}
\usepackage{array}
\usepackage{color}
\usepackage{amsmath}
\usepackage{amsxtra}
\usepackage{amstext}
\usepackage{amssymb}
\usepackage{latexsym}
\usepackage{dsfont}
\usepackage{subcaption}
\usepackage{lipsum}
\usepackage[colorlinks=true,allcolors=blue]{hyperref}
\usepackage[margin=1in]{geometry}

\newcommand{\bb}[1]{\mathbf{#1}}

\newcommand{\tm}[1]{\text{#1}}
\newcommand{\brak}[2]{\left\langle{#1}|{#2}\right\rangle}

\newcommand{\bA}{\mathbf{A}}

\newcommand{\bB}{\mathbf{B}}

\newcommand{\bE}{\mathbf{E}}
\newcommand{\bR}{\mathbf{R}}
\newcommand{\bx}{\mathbf{x}}
\newcommand{\bk}{\mathbf{k}}
\newcommand{\bq}{\mathbf{q}}

\newcommand{\bn}{\mathbf{n}}

\newcommand{\omegaa}{\omega_\alpha}
\newcommand{\omegaap}{\omega_{\alpha'}}

\newcommand{\cV}{\mathcal{V}}

\newcommand{\cF}{\mathcal{F}}
\newcommand{\cC}{\mathcal{C}_{\mathbf{k}\alpha}}
\newcommand{\cCp}{\mathcal{C}_{\mathbf{k}\alpha'}}

\newcommand{\brho}{\boldsymbol{\rho}}
\newcommand{\bN}{\boldsymbol{\nabla}}

\begin{document}
\title{Vacuum radiation and frequency-mixing \\ in linear light-matter systems}
\author{Niclas Westerberg}
\email{nkw2@hw.ac.uk}
\affiliation{Institute of Photonics and Quantum Sciences, Heriot-Watt University, EH14 4AS Edinburgh, United Kingdom}
\affiliation{School of Physics and Astronomy, University of Glasgow, Glasgow G12 8QQ, United Kingdom}
\author{Angus Prain}
\affiliation{School of Physics and Astronomy, University of Glasgow, Glasgow G12 8QQ, United Kingdom}
\affiliation{Institute of Photonics and Quantum Sciences, Heriot-Watt University, EH14 4AS Edinburgh, United Kingdom}
\author{Daniele Faccio}
\affiliation{School of Physics and Astronomy, University of Glasgow, Glasgow G12 8QQ, United Kingdom}
\affiliation{Institute of Photonics and Quantum Sciences, Heriot-Watt University, EH14 4AS Edinburgh, United Kingdom}
\author{Patrik \"Ohberg}
\affiliation{Institute of Photonics and Quantum Sciences, Heriot-Watt University, EH14 4AS Edinburgh, United Kingdom}

\begin{abstract}
Recent progress in photonics has led to a renewed interest in time-varying media that change on timescales comparable to the optical wave oscillation time. However, these studies typically overlook the role of material dispersion that will necessarily imply a delayed temporal response or, stated alternatively, a memory effect. We investigate the influence of the medium memory on a specific effect, i.e. the excitation of quantum vacuum radiation due to the temporal modulation. We construct a framework which reduces the problem to single-particle quantum mechanics, which we then use to study the quantum vacuum radiation. We find that the {delayed temporal response} changes the vacuum emission properties drastically: Frequencies mix, something typically associated with nonlinear processes, despite the system being completely linear. Indeed, this effect is related to the parametric resonances of the light-matter system, and to the parametric driving of the system by frequencies present locally in the drive but not in its spectrum.
\end{abstract}

\date{\today}
\maketitle

\section{Introduction}
Light experiences dispersion as it passes through an optical medium, such as the glass in your window or the water in your glass, and different frequencies appear to be travelling at different rates. On a quantum level, the vacuum inside the glass is different from the vacuum outside it. In light of recent studies that return to the problem of time-dependent media \cite{timerefrac1,timerefrac3,timerefracBook,timerefrac4,timerefrac2,AngusLiberati,AngusNic,shalaev}, it is worth asking if dispersion plays an additional, non-trivial, role also in a medium whose properties changes with time. 

Optical dispersion is of course well known, and is accurately described by the theory of macroscopic electrodynamics \cite{landau,jackson}, in which one ignores the microscopic make-up of the medium, replacing the chain of absorption and re-emission processes of the constituents (from which the dispersion originates) with a \textit{phenomenological} frequency-dependent permittivity $\varepsilon$. This greatly simplifies the problem on a classical level, but introduces some difficulties when attempting to quantise the theory, as the Lagrangian and Hamiltonian of the theory becomes ill-defined. Consequently, many different approaches have been pursued (a good review of which can be found in Refs.~\cite{MacroQEDOverview1,MacroQEDOverview2} and references therein). On a conceptual level these issues have now largely been resolved by introducing phenomenological microscopic degrees of freedom, often in the manner proposed by \citet{hopfield}, see for instance the work of \citet{barnett} or \citet{philbin}. {Such phenomenological microscopic degrees of freedom usually consists of a simplified version of medium constituent dynamics (i.e. microscopic detail), which nonetheless gives the same result at a macroscopic scale.} The presence of the medium directly affects the electromagnetic vacuum, leading to Casimir-Polder forces, as is discussed thoroughly in Ref.~\cite{forcesOnVacuum}. Calculations can however become complex within these frameworks, especially when introducing time-dependencies to the medium. 

{Time-dependent optical media have been the subject of a range of studies, examples of which include acceleration radiation in a plasma \cite{timerefrac1}, time-refraction \cite{timerefrac2,timerefrac3}, and more generically parametric oscillators \cite{arnol2013mathematical,paraComb1,paraComb2,paraMech1,paraMech2,paraFluid1,paraFluid2,paraElec1,paraElec2,paraLM2} and vacuum radiation \cite{AngusLiberati,AngusNic,referee1ref1,forcedVacuum}}. {It is also closely linked to polariton physics, that is, the physics of collective excitations of light-matter systems. These coupled systems offers rich physics that likewise has a rich history \cite{polaritonsOld,quantumPolaritons}. To name a few, it has been explored in the context of photon fluids in microcavities \cite{polCond,fluidsOfLightCarusotto}, cavity quantum electrodynamics \cite{cavityQED,cavityQED1,paraLM1}, optomechanical systems \cite{optoMech,optoMech1}, as well as surface-plasmon polaritons \cite{surfacePlasmonReview}. In this picture, the study of quantum vacuum radiation in time-dependent media becomes the study of polaritons excited from the vacuum state. This has natural links to excitations in temporally modulated quantum systems \cite{quantumModulationReview}, the dynamical Casimir effect \cite{DCEreview,circuitDCE,gjohansson} and quantum field theory on time-dependent backgrounds \cite{timeQFT1,birrelDavies,jacobson,uweCosmo,cosmoVisser,cosmoAngus,westerbergNJP}. Many studies have focused on microcavities and exciton-polaritons, see for instance Refs.~\cite{referee1ref2,referee1ref3,referee1ref4,referee1ref5,referee1ref6,referee1ref7,excitonPolariton,ultraStrongPolaritons,microscopicCavityDynCasi}, where strong light-matter coupling is possible. However, the phenomenon is not restricted to the strong-coupling limit, and bulk media was studied in for example Refs.~\cite{referee1ref1,belgiorno}, along with surface plasmon-polaritons in Refs.~\cite{vacuumSurfacePlasmon,DCEsurfacePlasmon}.}

{In this work we will study the temporal modulation of bulk media, or fibre-like scenarios, at multiple frequencies. Specifically, our aim is to develop models of quantum vacuum radiation relevant for experiments such as Refs.~\cite{enzD2} and \cite{modFibre}, due to recent progress in photonics as well as experimental interest. In particular, we examine the role of dispersion, especially with regards to the temporally delayed response, in the production of photons from the vacuum state due to the medium time-dependence. Interestingly, different physics transpire depending on whether it is the light-matter coupling (such as Rabi frequency) or the resonance frequency that is temporally modulated.} As we will discuss, the former acts similarly to direct driving whereas the latter, on which we will focus, is a type of parametric driving. Whilst the spectrum of vacuum radiation is qualitatively similar in both cases to first order in the size of the refractive index modulation, this is no longer the case at higher order{s} (also discussed in Ref.~\cite{angusSuperoscillations}). {We will therefore focus our attention on non-trivial second order effects, which is a topic of increasing importance with the advent of so-called epsilon-near-zero materials \cite{enz1,enz2,enz3,enzD1,enzD2} where changes to the refractive index in time can be in the order of unity.} We will employ a model for macroscopic electromagnetism where the microscopic degrees of freedom is treated phenomenologically, in the spirit of Hopfield \cite{hopfield}, and similar to Refs.~\cite{barnett, hansson, unruh, belgiorno}. This model allows to fully account for dispersion and memory effects. {As a result we uncover a frequency mixing mechanism that modifies the spectrum of the emitted photon pairs. 

Usually quantum vacuum radiation is emitted when the sum of two polariton frequencies match the frequencies contained within the spectrum of the modulation \cite{belgiorno,referee1ref3}. In our case, we modulate the resonance frequency at $\nu_1$ and $\nu_2$, and the spectrum is thus strongly peaked around these frequencies. However, multiple-frequency modulations will form an interference pattern in the time domain, which oscillates at frequencies outside the spectrum. Question then becomes whether or not energy can be absorbed by this. Interestingly, we find that frequency-mixed photons appear when the sum of two light-matter quasiparticle frequencies match $\nu_ 1$, $\nu_2$ or $|\nu_1\pm\nu_2|$. The latter are indeed the beating frequencies.} This not only provides a physical manifestation of time-dependent media but also provides an additional route for the detection of photons in a background free environment (i.e. at frequencies that are displaced from those of any input fields). Whilst frequency mixing is usually connected to nonlinear processes, here the underlying assumption is that the medium response is at all times linear. Instead, the mixing phenomenon is related to a parametric response of a coupled system. {In particular, we find that energy can be absorbed from the modulation interference pattern precisely because of the time-delayed response of the medium. In this process, energy is absorbed from the wave oscillating at $\nu_1$, stored until a (anti)quanta of energy is absorbed by the second wave oscillating at $(-)\nu_2$ (or vice-versa). The total energy $\left|\nu_1\pm\nu_2\right|$ is then emitted in the form of a polariton pair.} The latter is related to the `superoscillations' studied in Ref.~\cite{angusSuperoscillations}{, and the `bichromatic' driving briefly mention in Ref.~\cite{referee1ref6} can be seen as a special case of this.}

The manuscript is structured as follows: In Section~\ref{sec:model}, we define a microscopic phenomenological action for the light-matter system, whose classical equation of motion results in a common type of dispersion relation. We then define polariton branches and quantise using a path integral formalism in Section~\ref{sec:quantisation}. Transition amplitudes for temporally modulated media are then discussed in Section~\ref{sec:transAmp}, an in-depth example of which we treat in Section~\ref{sec:dispMix}. Discussion of the methods and concluding remarks are then presented in Section~\ref{sec:conclusion}. 

\section{The Model and effective action}\label{sec:model}
{It is well-known that the dispersive response of the medium complicates calculations. The origin of this complexity is the two distinct types of time dynamics at interplay: Optical parameters that change with time, as well as the time-delayed response of the medium. The time-delayed response is directly connected to dispersion, as the rate at which the medium constituents absorb and re-emit light depends on the frequency. Such frequency-dependence of the response implies, by necessity, that the Hamiltonian/Lagrangian is nonlocal in time. The medium is therefore characterised, in the time-domain, by a memory kernel connecting past events with the present \cite{landau,jackson,boyd}. In the context of macroscopic electromagnetism, a time-dependent medium is introduced by allowing a model parameter, such as the resonance frequency, to change with time. The resulting time-dependent permittivity is then described by a memory kernel which changes non-trivially with time.}

{In this work, we will model the optical medium as a set of harmonic oscillators $\mathbf{R}_i$ with natural oscillation frequencies $\Omega_i$ respectively, at a spatial density of $\rho$. As we will see shortly, these oscillation frequencies will act as the resonance frequencies of the medium. Note, we will use units such that $c = \hbar = \epsilon_0 = 1$ for notational simplicity. Coupling this to electromagnetism by dipole terms, quantified by dipolar coupling strengths $q_i$, yields the action
\begin{align}\label{eq:action0}
S_\gamma &= \int_{t_i}^{t_f} dt \int d^3x \;  \frac{1}{2}\left[\mathbf{E}^2-\mathbf{B}^2\right] \nonumber \\
S_{R} &= \sum_i \int_{t_i}^{t_f} dt \int d^3x \; \frac{\rho}{2}\left[\dot{\bR}_i^2-\Omega_i^2(\bb{x},t)\bR_i^2\right] \nonumber \\
S_\tm{int} &= \sum_i \int_{t_i}^{t_f} dt \int d^3x \; \left(\rho q_i\right) \; \mathbf{E}\cdot\bR_i,
\end{align}
where the natural oscillation frequencies $\Omega_i^2(\bb{x},t)$ can in general be space- and time-dependent. Now, the electric field is given by $\bb{E}~=~-\left(\partial_t\bA~+~\bN\varphi\right)$, where $\bA$ and $\varphi$ are the vector and scalar potentials respectively. In Coulomb gauge, i.e. when $\bN\cdot \bA = 0$, the equation of motion for the scalar potential is
\begin{equation}\label{eq:scalarEoM}
\nabla^2\varphi - \sum_i\rho q_i \bN\cdot\bb{R}_i = 0,
\end{equation}
which implies that $\bN\varphi = \sum_i\rho q_i \bR_i$. Starting with the Lagrangian density, defined as $S = \int dt \mathcal{L}$, for electromagnetism and the light-matter coupling, 
\begin{align*}
\mathcal{L}_{\gamma+\tm{int}} &=  \frac{1}{2}\left[\bE^2-\bB^2\right] + \sum_i \; \left(\rho q_i\right) \; \bE\cdot\bR_i,
\end{align*}
and substituting in $\varphi$ from Eq.~\eqref{eq:scalarEoM} leads to 
\begin{align*}
\mathcal{L}_{\gamma+\tm{int}} = \frac{1}{2}&\left[\dot{\bA}^2-\left(\bN\times\bA\right)^2\right] - \sum_i \; \left(\rho q_i\right) \; \dot{\bA}\cdot\bR_i \nonumber \\
&- \frac{1}{2}\sum_{ij} \; \left(\rho q_i q_j\right) \; \bR_i\cdot\bR_j.
\end{align*}
Since the last term is just quadratic in the oscillator fields $\bR_i$, we can re-diagonalise. Furthermore, this can be done without impacting the form of the action, since the oscillator parameters $\rho$, $\Omega_i$ and $q_i$ are all phenomenological, i.e. chosen to fit experimental data. 

We therefore arrive at an action describing the electromagnetic vector potential $\bb{A}$ coupled to a set of oscillators $\bb{R}_i$ by a dipole term, where the latter phenomenologically take into account the microscopic details of the matter degree of freedom, given by}
\begin{align}\label{eq:action}
S_\gamma &= \int_{t_i}^{t_f} dt \int d^3x \;  \frac{1}{2}\left[\dot{\bA}^2-\left(\bN\times\bA\right)^2\right] \nonumber \\
S_{R} &= \sum_i \int_{t_i}^{t_f} dt \int d^3x \; \frac{\rho}{2}\left[\dot{\bR}_i^2-\Omega_i^2(\bb{x},t)\bR_i^2\right] \nonumber \\
S_\tm{int} &= \sum_i \int_{t_i}^{t_f} dt \int d^3x \; \left(-\rho q_i\right) \; \dot{\bA}\cdot\bR_i,
\end{align}
{where $\bb{R}_i$ is the position of each oscillator in its potential well and $\rho$ is the density of oscillators.} This action is inspired by the Hopfield models employed in Refs.~\cite{hopfield, barnett, hansson, unruh, belgiorno}. In the case of constant $\Omega_i(\bx,t) \equiv \Omega_i$, we find that Eq.~\eqref{eq:action} leads to a dispersion relation for the electric field of the familiar Sellmeier form,
\begin{equation}\label{eq:homdisp}
D(\bk,\omega) \equiv -|\bk|^2 + \omega^2\left(1-\sum_i\frac{g_i^2}{\omega^2-\Omega_i^2}\right) = 0,
\end{equation}
where $g_i^2 = \rho q_i^2$ are the effective plasma frequencies of the medium resonances. This corresponds to a refractive index {
\begin{align*}
n^2(\omega) = 1-\sum_i \frac{g_i^2}{\omega^2-\Omega_i^2},
\end{align*}
as is widely adopted in the optics literature \cite{hecht}}. In other words, the above action is a suitable starting point for modelling any dielectric where absorption is negligible. {From this we can see that a time-dependent $\Omega_i$ induces temporal changes in the refractive index. We note that it is however also possible to create a time-dependent medium through a coupling strength $q_i$ that depends on time. This has been studied in various scenarios, referred to as a time-dependent Rabi frequency, and we will not delve deeply into this scenario here. Already at this stage we can see, from the above action [Eq.~\eqref{eq:action}], that such a time-dependence will act more akin to a direct driving force than a parametric drive.}

Similarly to Ref.~\cite{hansson}, we want to compute an effective action for the photons, $S_\tm{eff}[\bA]$, by integrating out the oscillator degree of freedom. Schematically, we do this by computing
\begin{equation*}
\exp i S_\tm{eff}[\bA] = \mathcal{N}\int \mathcal{D}\bR_i \; e^{i\left(S_\gamma[\bA] + S_R[\bR_i] + S_\tm{int}[\bA,\bR_i] \right) },
\end{equation*}
with the boundary conditions $\bR_i(\bx,t_i) = \bR_i(\bx,t_f) = 0$, as we are not interested in the dynamics of $\bR_i$. {In this path integral we integrate over each possible configuration of the oscillator position $\mathbf{R}_i$ as a function of time that fulfils the stated boundary conditions, as defined in Ref.~\cite{feynman}.} However, as the coupling in $S_\tm{int}[\bA,\bR_i]$ is {linear}, it is easy to show that the quantum fluctuation of $\bR_i$ does not affect $\bA$ and is contained in the normalisation constant $\mathcal{N}$ (here set to unity) \cite{feynman}. Therefore, performing this path integral for $\bR_i$ with the above boundary conditions is equivalent to solving the classical equation of motions for $\bR_i$ driven by $q_i \dot{\bA}$ \cite{handbook}. This can be done by method of Green's functions, that is, by solving 
\begin{equation}\label{eq:propagator}
\left[\partial_t^2+\Omega^2_i(\bx,t)\right]\Delta_i = -\delta(t-t'),
\end{equation}
with boundary conditions $\Delta_i(\bx,t_f,t') = \Delta_i(\bx,t_i,t') = 0$. We can link this to usual optics parameters by noting that the medium response function, commonly denoted as $\chi(\bx,t,t')$ \cite{boyd}, is given by $\chi(\bx,t,t')=\sum_i\partial^2_t\Delta_i(\bx,t,t')$. In other words, substituting 
\begin{equation*}
\bR_i(\bx,t)\rightarrow \frac{q_i}{2}\int_{t_i}^{t}dt'\Delta_i(\bx,t,t')\dot{\bA}(\bx,t')
\end{equation*}
into $S_\tm{int}[\bA,\bR]$ yields the effective action for photons. In order to make this more tangible, let us also expand the vector potential in the polarisation vectors $\bA~=~\sum_{\lambda=1,2} \mathbf{e}_\lambda A_\lambda$, where $\mathbf{e}_{\lambda} \cdot \mathbf{e}_{\lambda'}~=~\delta_{\lambda,\lambda'}$ defined with respect to some reference vector $\mathbf{p}$ such that $\mathbf{e}_{\lambda}\cdot\mathbf{p}=0$. We should note here that since $\bA$ is completely transverse, so are the oscillators $\bR_i$, and thus satisfy the Coulomb `gauge' condition $\bN\cdot\bR_i = 0$. Through this, we find the effective action 
\begin{align}\label{eq:effAc}
S_\tm{eff}[\bA] = \sum_\lambda  \frac{1}{2}\bigg(\int_{t_i}^{t_f} dt &\int d^3x \; \bigg[\dot{A_\lambda}^2 - \left(\bN A_\lambda\right)^2 \nonumber\\
 &- \sum_i g_i^2 \int_{t_i}^{t_f}dt' \; \dot{A_\lambda}(\bx,t)\Delta_i(\bx,t,t')\dot{A_\lambda}(\bx,t')  \bigg]\bigg),
\end{align}
where $\Delta_i$ is the oscillator propagator given in Eq.~\eqref{eq:propagator}{, with $g_i^2 = \rho q_i^2$ being the effective plasma frequencies for each resonance respectively.}

Since the two polarisations de-couple, we will from here on drop the $\lambda$ subscript for notational simplicity, and work only with the scalar quantity $A(\bx,t)$. This is so far general, and we have specified neither the space nor the time-dependence of $\Omega^2_i(\bx,t)$. In the next section, we will consider the case of a static but inhomogeneous set of oscillators, such that $\Omega^2_i(\bx,t)~\equiv~\Omega_i^2(\bx)$. The spatial dependence will be taken into account by expanding in an appropriate set of normal modes $u_\bk(\bx)$, the exact form of which depends on the physical situation. 

{Let us start with the equation of motion for the vector potential from Eq.~\eqref{eq:effAc}, under the assumption that the oscillator frequency is time-independent. This is given by
\begin{equation*}
\left(\partial_t^2-\nabla^2\right)A(\bx,t)+\sum_i\int_{t_i}^{t_f}dt'\dot{\Delta}_i(\bx,t,t')\dot{A}(\bx,t')=0.
\end{equation*}
The goal is now to expand the vector potential in a set of normal modes 
\begin{equation*}
A(\bx,t) = \sum_\bk u_\bk(\bx)\int \frac{d\omega}{2\pi}A_\bk(\omega)e^{-i\omega t}
\end{equation*}
defined such that $\int d^3x \; u^*_{\bk}(\bx)u_{\mathbf{l}}(\bx) = \delta_{\bk\mathbf{l}}$. The form of the mode functions $u_k(\bx)$ depends on the physical scenario. We thus look for functions that satisfy 
\begin{equation}\label{eq:modefunc}
\left[\nabla^2 +\omega^2\left(1-\sum_i\frac{g_i^2}{\omega^2-\Omega_i^2(\bx)}\right)\right]u_\bk(\bx)=0,
\end{equation}
with appropriate boundary conditions for the situation. In this work, we will focus on bulk media, and as such there is no spatial dependence on the oscillator frequency ($\Omega_i(\bx)\equiv \Omega_i$). A brief aside into a fibre-like scenario can be found in Appendix~\ref{app:normalmodes}. There are nonetheless multiple ways of expanding in terms of normal modes for bulk media.

\subsection{Plane waves}
For bulk media, a natural choice of normal modes are the momentum modes 
\begin{align*}
u_\bk(\bx) \propto \exp(i\bk\cdot\bx)
\end{align*}
Here we find the dispersion relation given by Eq.~\eqref{eq:homdisp}. 

\subsection{Paraxial waves}
In most experimental scenarios however, the simple plane waves are not accessible, and are instead replaced by structured paraxial beams. Let us once again consider a homogeneous bulk medium where $\Omega^2_i(\bx)~\equiv~\Omega_i^2$, but where we restrict Eq.~\eqref{eq:modefunc} to the paraxial limit, with the $z$-direction chosen to be the propagation direction. 

In other words, let $u_\bk(\bx) = u_k(\brho,z)e^{ikz}$ with $k$ being the momentum in the $z$-direction and $q^2/2k^2 \ll 1$, where $\brho$ is the transverse plane coordinates and $\bq$ its associated momentum. Similarly to Ref.~\cite{OAMQFT}, we then find that $k$ must follow the dispersion relation of Eq.~\eqref{eq:homdisp}, and $u_k(\brho)$ satisfies the paraxial wave equation
\begin{equation*}
\left(\nabla^2_\perp + 2ik\partial_z\right) u_k(\brho) = 0,
\end{equation*}
where $\nabla^2_\perp$ is the transverse Laplacian. Solutions include the familiar Laguerre-Gaussian modes and Hermite-Gaussian modes \cite{GoodmanOptics}.
}
\section{Quantisation}\label{sec:quantisation}
\begin{figure*}
\includegraphics[width=0.9\textwidth]{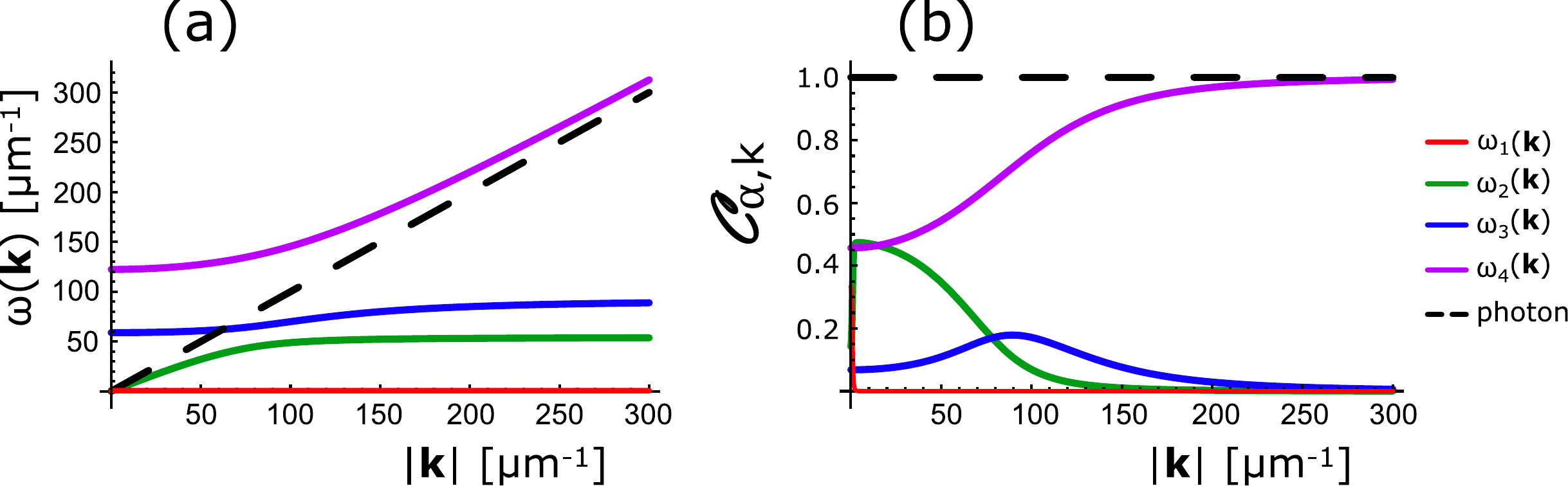}
\caption{\textbf{(a)} Polariton branches of fused silica (solid) in units of $c=1$, as well as free photon dispersion (dashed). Note that the optical regime lies between $3-20\mu\text{m}^{-1}$, whereas the majority of the structure in the spectrum is at higher energy. \textbf{(b)} Overlap coefficient $\mathcal{C}_{\bk\alpha}$ for the different branches. Note that the value varies between zero and unity depending on whether a specific branch $\alpha$ can be characterised as `photon-like' or not.}\label{fig:branches}
\end{figure*}

It is often noted that constructing a quantum field theory reduces to quantising an infinite set of harmonic oscillators \cite{feynman,loudon,srednicki}, one for each (commonly continuous) position/momentum. By expanding a field in terms of suitable normal modes however, one can take this seriously and proceed to quantise each normal mode. This is usually done within the canonical quantisation scheme, but we will here use a path integral language. Whilst this quantisation technique is known for plane waves in vacuum \cite{feynman,rovelli}, it is not commonly employed for computations, nor has it been generalised for dispersion. As we shall show here however, we find this technique particularly suitable for tackling the type of problems addressed by macroscopic quantum electrodynamics. 

Let us start by re-writing the effective action of Eq.~\eqref{eq:effAc} in the frequency domain, yielding
\begin{align*}
S_\tm{eff}= \frac{1}{2}&\int\frac{d\omega}{2\pi} \int d^3x \; A^*(\omega,\bx)\bigg[\nabla^2 + \;\omega^2\left(1-\sum_i\frac{g_i^2}{\omega^2-\Omega_i^2(\bx)}\right)\bigg]A(\omega,\bx),
\end{align*}
where $A^*(\omega,\bx) = A(-\omega,\bx)$ as $A(t,\bx)$ is a real quantity. Here we integrated by parts on the $\bN$-term and used that $\Delta_i(\bx,t,t')$ is diagonal in the frequency domain when $\Omega_i$ is time-independent. We now expand this in terms of normal modes, defined such that $\int d^3x \; u^*_{\bk}(\bx)u_{\mathbf{l}}(\bx) = \delta_{\bk\mathbf{l}}$, finding 
\begin{align}\label{eq:effAc1}
S_\tm{eff}= \frac{1}{2}\sum_{\bk}\int\frac{d\omega}{2\pi} \; A_\bk^*(\omega)D(\bk,\omega)A_\bk(\omega),
\end{align}
where $A_\bk^*(\omega) = A_{-\bk}(-\omega)$ and $D(\bk,\omega)$ depends on the particular normal modes used (see Appendix~\ref{app:normalmodes}). As an example, for plane waves this reduces to Eq.~\eqref{eq:homdisp}.

The solution of the classical equations of motion for each normal mode takes the form of $A_\bk \propto \exp(\pm i\omega_\alpha t)$, where $\omega_\alpha(\bk)$ is given by the poles of $1/D(\bk,\omega)$. This defines the quasiparticles of the system. In other words, by solving $D(\bk,\omega) = 0$ for $\omega$ as a function of normal mode label $\bk$, we find $N$ quasiparticle branches. These are usually referred to as \textit{polaritons}. The exact number of polariton branches depends on the explicit form of $D(\bk,\omega)$. We will label these branches by the subscript $\alpha$, and an example can be seen in Fig.~\ref{fig:branches}(a). Inspired by Ref.~\cite{hansson}, we can do the following field transformation
\begin{align}\label{eq:fieldTrans}
A_{\bk}(\omega) &= \sqrt{\frac{\omega^2-\omega^2_\alpha(\bk)}{D(\bk,\omega)}}A_{\bk\alpha}(\omega) \nonumber\\
&= \mathcal{P}_{\bk\alpha}(\omega) A_{\bk\alpha}(\omega)
\end{align}
in order to define a polariton action. Note, this transformation is always well-defined as the $1/D(\bk,\omega)$ diverges at the same points and at the same rate as $\omega^2-\omega^2_\alpha(\bk)$ goes to zero. The action of Eq.~\eqref{eq:effAc1} is nonlocal in time (i.e. dispersive) in field-coordinates $A_\bk(\omega)$: By this field-transformation, we trade nonlocality in time for nonlocality in space. {This simplifies the quantisation procedure whilst still taking dispersion into account, as dispersion is now implicit in the definition of the polariton fields and their dependence on the momentum mode $\bk$. Temporal nonlocalities in quantum theory can be treated, but usually at a computational cost since one must now define a (commonly) infinite set of conjugate momenta (see for instance the discussions in Refs.~\cite{QTnonloc1,QTnonloc2}). A spatial nonlocality on the other hand, which here means that the polariton frequencies $\omegaa(\bk)$ contain terms of higher order than $k^2$, is straightforward to tackle since we will treat each momentum mode $\bk$ independently.} 

Written in field-coordinates $A_\alpha$, and after transforming back into temporal space, the action is that of a set of complex harmonic oscillator
\begin{align}\label{eq:effAc2}
S_\tm{eff}= \sum_{\bk,\alpha}\int_{t_i}^{t_f} dt\; \frac{1}{2}\left(|\dot{A}_{\bk\alpha}(t)|^2-\omega^2_\alpha |A_{\bk\alpha}(t)|^2\right).
\end{align}
This is the action with which we will work. From now on, we will be working with the dynamics of single normal modes $\bk$, so we will drop the sum over $\bk$ and corresponding identifier in order to simplify notation. Also, we will drop the index $\alpha$ on all but the mode frequency $\omegaa$ for the same reason.

Although this is a field theory, in terms of normal modes, all the usual techniques from single-particle quantum mechanics apply. This can be quantised in the manner most familiar to the reader. {In this work, we choose a path integral method as it allows for a straightforward definition of time-nonlocal perturbation theory.} 

As is usual in path integral quantisation, we want to add the driving terms $J A^*$ and $J^* A$ to the action for future use. These driving terms physically originate from free currents in the system, i.e. the movements of free charges. We will however not consider physical driving here, but use the driving terms for computational purposes. After simplifying the notation and adding the driving, the action takes the form
\begin{align}\label{eq:effAcSimp}
S_\tm{eff}\left[J,J^*\right]= \int_{t_i}^{t_f} dt\; \frac{1}{2}\left(|\dot{A}|^2-\omega^2_\alpha |A|^2 + J^*A + J A^*\right).
\end{align}
Let us now proceed by calculating the polariton transition amplitude
\begin{equation*}
\brak{A_f,t_f}{A_i,t_i}_J = \int \mathcal{D}A \mathcal{D}A^* \; \exp\left(i S_\tm{eff}\left[J,J^*\right] \right),
\end{equation*}
where we have the boundary condition $A(t_i) = A_i$ and $A(t_f) = A_f$. Here we calculate the probability amplitude for a polariton in branch $\alpha$, normal mode $\bk$ and polarisation $\lambda$, starting with field amplitude $A_i$ at time $t_i$ and transitioning to field amplitude $A_f$ at time $t_f$ \footnote{This can be computed in the same manner as for a simple harmonic oscillator \cite{feynman,rovelli}, though some extra care should be taken with the indices. Note that we do not use the boundary conditions $A_f = A_i = 0$ and the subsequent Lehmann-Symanzik-Zimmermann reduction formalism \cite{srednicki} here as we do not necessarily deal with scattering states.}.

First we note that the quantum fluctuations decouple from the classical dynamics, as the action in Eq.~\eqref{eq:effAcSimp} is quadratic in the fields. As a consequence, the transition amplitude factorises as $\mathcal{F}(T)\exp\left[iS_\text{cl}\right]$, {where $S_\text{cl} = S_\tm{eff}[A_\tm{cl}]$ is the classical action and the pre-factor $\mathcal{F}(T)$ is determined by the quantum fluctuations $\eta$. We here define the quantum fluctuation by splitting the field into classical and quantum components $A = A_\tm{cl} + \eta$, such that $\eta(t_i)=\eta(t_f) = 0$.} Explicitly, this pre-factor is given by
\begin{align*}
\mathcal{F}(T) = \int_{\eta(t_i)=0}^{\eta(t_f)=0} \mathcal{D}\eta\mathcal{D}\eta^* \; e^{\frac{i}{2} \int_{t_i}^{t_f} dt\; \left(|\dot{\eta}|^2-\omega^2_\alpha |\eta|^2\right)}.
\end{align*}  
We calculate the classical action using the equation of motion along with the boundary conditions at $t_i$ and $t_f$. Finally, we find the transition amplitude
\begin{equation}\label{eq:fieldProp}
\brak{A_f,t_f}{A_i,t_i}_J = \left(\frac{\omegaa}{4\pi i \sin \omegaa T}\right)e^{i S_\tm{cl}[J,J^*]},
\end{equation}
where $T=t_f-t_i$ and the classical action $S_\tm{cl}[J,J^*]$ is that of a complex driven simple harmonic oscillator. See Appendix~\ref{app:amplitude} for a detailed calculation. As usual, this expression contains all information required for computations.

\subsection{Connecting polaritons and photons}
{The field transformation in Eq.~\eqref{eq:fieldTrans} also has a physical interpretation. In doing this, we project the photon field in terms of polariton fields. The expansion coefficients, a generalisation of the co-called Hopfield coefficients \cite{hopfield}, are given by $\mathcal{P}_{\bk\alpha}(\omega)$. As we are always quadratic in the fields, it is convenient to define the squared coefficients $\mathcal{C}_{\bk\alpha} \equiv \mathcal{P}^2_{\bk\alpha}(\omega)$. These are given by
\begin{align*}
\mathcal{C}_{\bk\alpha} = \lim_{\omega\rightarrow\omegaa}\left[\frac{\omega^2-\omegaa^2}{D(\bk,\omega)}\right] = \frac{\prod_{i}\left(\omegaa^2-\Omega^2_i\right)}{\prod_{\gamma \neq \alpha}\left(\omegaa^2-\omega^2_\gamma\right)}.
\end{align*}
Here we have used the fact that the polaritons live on-shell (i.e. $\omega~=~\omega_\alpha$), and that $D(\bk,\omega)\prod_{i}\left(\omega^2-\Omega^2_i\right)=\prod_{\gamma}\left(\omega^2-\omega^2_\gamma\right)$. It can easily be shown that $0 \leq \mathcal{C}_{\bk\alpha} \leq 1$, and corresponds physically to a factor describing the degree to which the polariton is `photon-like'. In other words, in spectral regions where $\omegaa \simeq k$, this factor is close to unity, and \textit{vice versa}. In Fig.~\ref{fig:branches}(b), an example of this can be seen.

We should note that in order to go from polariton observables to photon observables, the field transformation in Eq.~\eqref{eq:fieldTrans} needs to be undone. In general, integral expressions will come with factors of $\mathcal{C}_{\bk\alpha}$ when transforming from polariton to photon degrees of freedom, although in the actual path integral it can be absorbed into the normalisation.}

\section{Transition amplitudes}\label{sec:transAmp}
A time-dependent medium can generally change the number of polaritons in the system: Quanta can be excited from the vacuum \cite{birrelDavies}, whose accompanied spectrum is of interest, and like-wise polaritons can be absorbed into the vacuum. The former is the vacuum radiation. Each process has a transition amplitude $G^{\bk\alpha}_{mn \leftarrow pq}(t_f,t_i)$, denoting a transition from a $(pq)$-state with $p+q$ polaritons at time $t_i$ into a $(mn)$-state with $m+n$ polaritons at time $t_f$, whose absolute square gives the associated probability. Here we will first consider this general situation. We will once again drop the $\bk$ identifier to simplify notation, unless otherwise stated. Throughout this, we will use a quantisation box of volume $\cV$, as is standard (see Ref.~\cite{loudon}), and the normal modes used take the form 
\begin{align*}
u_\bk(\bx) = e^{i\bk\cdot\bx}/\sqrt{\cV}.
\end{align*}
Also, we should note that these transition amplitudes are the polariton Fock space propagators. However, we will first take a detour into a system where driving is present, as this links directly to a time-dependent medium in a perturbative setting.

\subsection{Generating functionals}
Let us first consider a driven medium, whose amplitudes will act later as generating functionals when considering time-dependent media perturbatively. We will first calculate vacuum persistence amplitude $G^{J}_{00\leftarrow 00}(t_f,t_i)$, denoted $G^{J}_{00}(t_f,t_i)$ for notational simplicity. This is given by the Gaussian integral
\begin{align*}
G^{J}_{00}(t_f,t_i) = &\int d^2A_{f}d^2A_{i} \; \Psi^*_{00}(A_{f})\brak{A_f,t_f}{A_i,t_i}_J\Psi_{00}(A_i),
\end{align*}
where $d^2A = dAdA^*$ and $\Psi_{00}(A)$ is the time-independent version of the groundstate wavefunction seen in Appendix~\ref{app:wavefunctions}. Note that in this $A$ is a complex variable, and not a function. Computing this yields
\begin{align*}
G^{J}_{00}(t_f,t_i) = \exp\left[-\frac{1}{4\omegaa}\int dt \int dt' J(t)\cos\omegaa(t-t')J^*(t') \right]e^{-i\omegaa T}.
\end{align*}
As can be expected, this is simply the generalisation of the vacuum persistence amplitude in Ref.~\cite{feynman} to the case of a complex harmonic oscillator. 

However, this calculation becomes increasingly complex for higher energy states, and we will therefore use a trick similar to what is done in Appendix~\ref{app:wavefunctions} in order to derive the Fock wavefunctionals $\Psi_{mn}$. That is, we use the wavefunctionals 
\begin{equation*}
\phi_a(A) = \sqrt{\frac{\omegaa}{2\pi}}e^{-\omegaa\left|A-a\right|^2/2},
\end{equation*}
and calculate the transition amplitude 
\begin{align}\label{eq:genInt}
F(b,a)_J = &\int d^2A_{f}d^2A_{i}\;\phi_b^*(A_{f})\brak{A_f,t_f}{A_i,t_i}_J\phi_a(A_{i}).
\end{align}
This amplitude can be seen as a generating functional of sort. If we expand $\phi_a$ and $\phi_b$ in terms of the Fock wavefunctionals $\Psi_{mn}$, we find that 
\begin{align*}
F(b,a)_J &= \sum_{mnpq} \psi^*_{mn}(b)\psi_{pq}(a) \int d^2A_{f}d^2A_{i}\;\Psi_{mn}^*(A_{f})\brak{A_f,t_f}{A_i,t_i}_J\Psi_{pq}(A_{i}) \nonumber \\
&= \sum_{mnpq} \psi^*_{mn}(b)\psi_{pq}(a) G^J_{mn\leftarrow pq}(t_f,t_i).
\end{align*}
In this way, we find the transition amplitudes
\begin{align*}
G^J_{mn\leftarrow pq}(t_f,t_i) = G^\alpha_{00}&\frac{(-1)^{m+q}e^{-i E_{mn} T}}{\sqrt{m!n!p!q!}}H_{nq}(i\beta_+,-i\beta_-^*)H_{pm}(i\beta_-,-i\beta_+^*)\nonumber
\end{align*}
where $E_{mn} = (m+n)\omegaa$, $H_{mn}$ are the complex Hermite polynomials given in Eq.~\eqref{eq:hermite} (also Refs.~\cite{complexHermite1,complexHermite2,complexHermite3}), and
\begin{equation*}
\beta_{\pm}^{(*)} = \frac{1}{\sqrt{4\omegaa}}\int_{t_i}^{t_f}dt \; e^{\pm i\omegaa (t-t_i)}J^{(*)}(t).
\end{equation*}
The explicit form of $F(b,a)_J$ can be found in Appendix~\ref{app:transition}. This captures all processes possible.

We can be a bit more explicit and ask ourselves what is the amplitude of exciting two polaritons back-to-back from the vacuum into mode $\bk$ in branch $\alpha$:
\begin{align}\label{eq:drivenAmp}
G^{J}_{11\leftarrow 00} &= \left(\frac{1}{4\omegaa}\right) \int dt \int dt' \; J(t)e^{i\omegaa t}e^{i\omegaa t'}J^*(t')\exp\left[-\frac{1}{4\omegaa}\int dt \int dt' J(t)\cos\omegaa(t-t')J^*(t')\right],
\end{align}
where we have ignored global phases.

\begin{figure}
\includegraphics[width=0.35\textwidth]{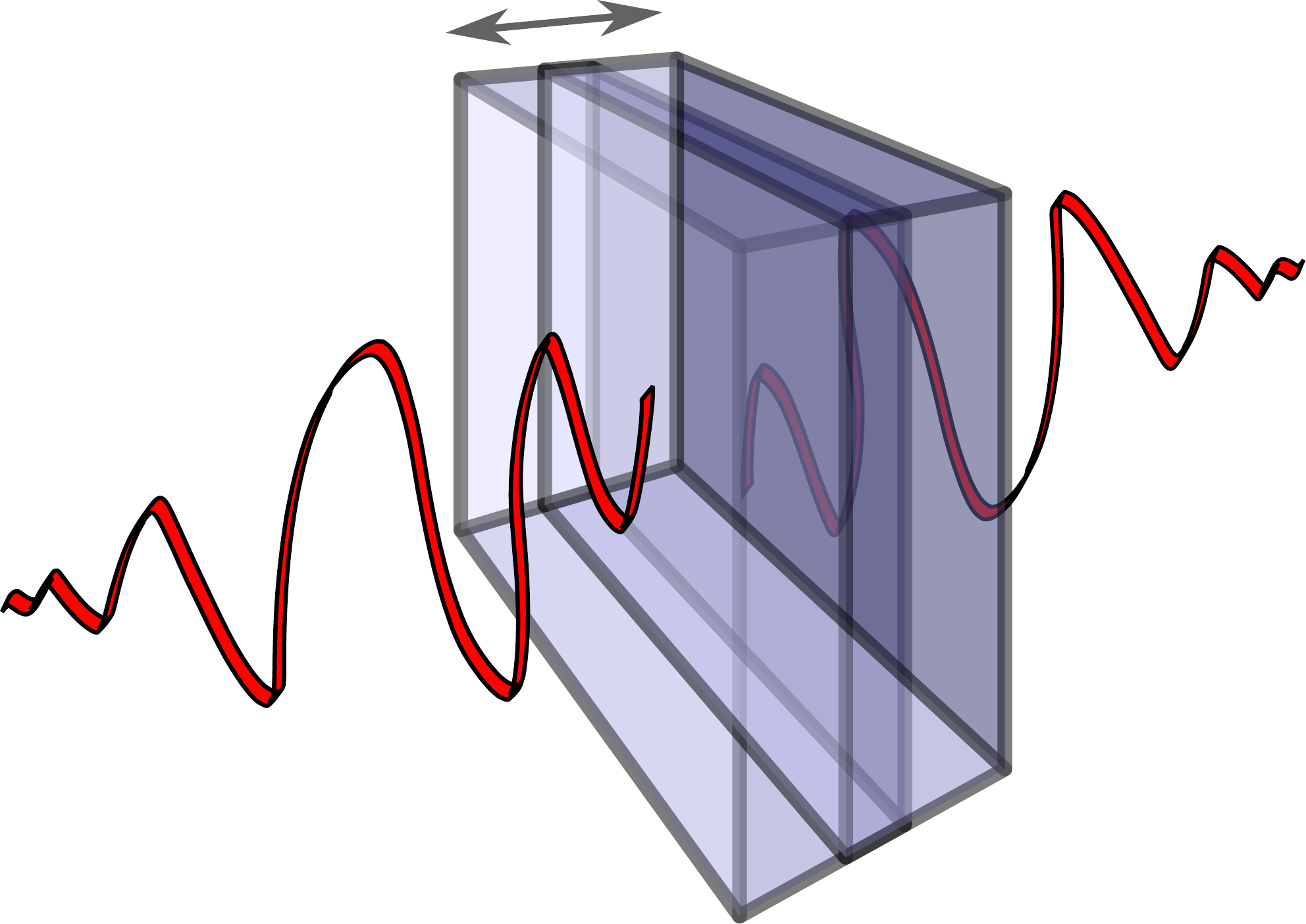}
\caption{Schematic of a time-dependent medium, and back-to-back polaritons generated thereof. The size of the medium represents the instanteneous permittivity.}\label{fig:sketches}
\end{figure}

\subsection{Time-dependent medium}
Let us now turn our attention to time-dependent media. In particular, let us consider a homogeneous medium with a weak space- and {time-dependent resonance frequencies,} i.e. $\Omega^2_i(\bx,t)\equiv \Omega^2_i\left[1+f_i(\bx,t)\right]$ where $|f| \ll 1$. We can then perturbatively construct the oscillator propagators (Eq.~\eqref{eq:propagator}) in orders of $|f|$: $\Delta_i(\bx,t,t') = \Delta^0_i(\bx,t,t') + \Delta^1_i(\bx,t,t') + \Delta^2_i(\bx,t,t') + \mathcal{O}(|f|^3)$. In the frequency domain, this leads to
\begin{align}\label{eq:oscprop}
\Delta^0_i(\bx,\omega,\omega') &= \frac{2\pi\delta(\omega+\omega')}{\omega^2-\Omega_i^2} \nonumber\\
\Delta^1_i(\bx,\omega,\omega') &= \Omega_i^2\frac{\tilde{f}_i(\bx,\omega+\omega')}{\left(\omega^2-\Omega_i^2\right)\left(\omega'^2-\Omega_i^2\right)} \nonumber\\
\Delta^2_i(\bx,\omega,\omega') &= \Omega_i^4 \int \frac{d\omega''}{2\pi} \frac{ \tilde{f}_i(\bx,\omega'')}{\left[(\omega-\omega'')^2-\Omega_i^2\right]}\frac{\tilde{f}_i(\bx,\omega+\omega'-\omega'')}{\left[\omega^2-\Omega_i^2\right]\left[\omega'^2-\Omega_i^2\right]}.
\end{align}
The $0^\tm{th}$-order is simply the usual propagator, leading to the dispersion relation in Eq.~\eqref{eq:homdisp}. {We can also relate the shift in oscillator frequency to the change in refractive index through 
\begin{align}\label{eq:deltan}
\delta n(\omega) = \sum_i \delta n_i(\omega) = \sum_i \left[-\frac{g_i^2\Omega_i^2}{\left(\omega^2-\Omega_i^2\right)^2}\left(\frac{\epsilon_i}{2 n(\omega)}\right)\right],
\end{align} 
}where $\epsilon_i$ is the characteristic amplitude of $f_i$. As we will see below, we can perform the same field transformation as before (Eq.~\eqref{eq:fieldTrans}) and we arrive at the polariton action in Eq.~\eqref{eq:effAc2}. The higher order propagators translate into (perturbative) potentials for the polaritons. {We cannot however trade the temporal nonlocality of the higher order potentials for additional spatial nonlocality, because these terms are not diagonal in frequency space (i.e. $\omega' \neq -\omega$). Therefore these will be temporally nonlocal \textit{two-time} potentials, also in the polariton field coordinates.} For clarity, let us proceed step-by-step. 

Before transforming to the polariton fields we have the effective action 
\begin{align*}
S_\tm{eff}= &\frac{1}{2}\sum_{\bk}\int\frac{d\omega}{2\pi} \;  A_\bk(\omega)D(\bk,\omega)A^*_\bk(\omega) \nonumber \\
&- \frac{1}{2}\sum_{\bk,\bk'}\int\frac{d\omega}{2\pi}\frac{d\omega'}{2\pi} \; A_\bk(\omega)\sigma_{\bk'-\bk}(-\omega,\omega')A^*_{\bk'}(\omega'),
\end{align*}
where we have let $\omega'\rightarrow -\omega'$ in the second integral, and we have defined the auxiliary propagator
\begin{equation}\label{eq:polprop}
\sigma_{\bk}(\omega,\omega')= \frac{1}{\cV}\sum_i g_i^2 \; \omega\omega'\left[\Delta^1_i(\bk,\omega,\omega')+\Delta^2_i(\bk,\omega,\omega')\right].
\end{equation}
Applying the polariton transformation in Eq.~\eqref{eq:fieldTrans} to the above action, and simplifying the notation, yields
\begin{align}\label{eq:effAc3}
S_\tm{eff}= &\int_{t_i}^{t_f} dt\;  \frac{1}{2}\left(|\dot{A}_{\bk\alpha}(t)|^2-\omega^2_\alpha |A_{\bk\alpha}(t)|^2\right) \nonumber \\
&- \frac{1}{2}\sum_{\bk'\alpha'} \int_{t_i}^{t_f} dt\int_{t_i}^{t_f} dt'\; A_{\bk\alpha}(t)\sigma^{\alpha\alpha'}_{\bk\bk'}(t,t')A_{\bk'\alpha'}^*(t'),
\end{align}
where we sum over momenta $\bk'$ and polariton branches $\alpha'$. Also, we have now defined the polariton projected auxiliary propagator $\sigma^{\alpha\alpha'}_{\bk\bk'}(t,t')$ as
\begin{align}\label{eq:polprop1}
\sigma^{\alpha\alpha'}_{\bk\bk'}(t,t') = \int&\frac{d\omega}{2\pi}\frac{d\omega'}{2\pi}e^{-i\omega t}e^{-i\omega' t'}\mathcal{P}_{\bk\alpha}(\omega) \sigma_{\bk'-\bk}(\omega,\omega')\mathcal{P}_{\bk'\alpha'}(\omega') ,
\end{align}
with $\mathcal{P}_{\bk\alpha}(\omega)$ being the polariton projection operator defined in Eq.~\eqref{eq:fieldTrans}. This takes the form of a complex harmonic oscillator, along with an additional \textit{two-time} harmonic potential
\begin{equation}\label{eq:pertPot}
V\left(A(t),A^*(t')\right) = \sum_{\bk'\alpha'}  A_{\bk\alpha}(t)\sigma^{\alpha\alpha'}_{\bk,\bk'}(t,t')A_{\bk'\alpha'}^*(t'),
\end{equation}
which connects the normal mode at $\bk$ with the one at $\bk'$. This latter term we will treat perturbatively, which is done by computing 
\begin{align}\label{eq:pertAmpCalc}
G^{V}_{mn\leftarrow pq}(t_f,t_i) = \exp\left(-\frac{i}{2}\sum_{\bk'\alpha'} \int_{t_i}^{t_f} dt\int_{t_i}^{t_f} dt'\; \sigma^{\alpha\alpha'}_{\bk,\bk'}(t,t')\frac{\delta}{i\delta J_{\bk\alpha}^*(t)}\frac{\delta}{i\delta J_{\bk'\alpha'}(t')}\right)G^{J}_{mn\leftarrow pq}(t_f,t_i)\bigg|_{J=0}.
\end{align}
There are two separate sectors here: Either polaritons are excited from the vacuum into the same polariton branch, or two separate ones. We will treat these two sectors separately for clarity, and will be referred to as intrabranch and interbranch vacuum radiation respectively. In both cases, we are interested in the probability amplitude of exciting a polariton pair back-to-back, as illustrated in Fig.~\ref{fig:sketches}.

\subsubsection{Intrabranch vacuum radiation}
Let us first consider the case when $\alpha'=\alpha$ in the perturbative potential of Eq.~\eqref{eq:pertPot}. We can then compute the necessary functional derivatives to Eq.~\eqref{eq:drivenAmp}. This yields 
\begin{align}\label{eq:timedepAmpFreq}
G^\text{intra}_{11 \leftarrow 00} = &\frac{i}{8\omegaa}\sigma^{\alpha\alpha}_{\bk\bk}(\omegaa,\omegaa) - \frac{1}{64\omegaa^2} \bigg[\sigma^{\alpha\alpha}_{\bk\bk}(\omegaa,\omegaa)\sigma^{\alpha\alpha}_{\bk\bk}(\omegaa,-\omegaa) + \sigma^{\alpha\alpha}_{\bk\bk}(\omegaa,\omegaa)\sigma^{\alpha\alpha}_{\bk\bk}(-\omegaa,\omegaa)\bigg] + \mathcal{O}(\sigma^3), 
\end{align}
where we have expanded to second order for consistency, and considered scattering states where $t_i = -\infty$ and $t_f = \infty$. We have also ignored the overall phases. Substituting the auxiliary propagator $\sigma^{\alpha\alpha}_{\bk\bk}$ in terms of oscillator propagators, Eqns.~\eqref{eq:polprop1} and \eqref{eq:polprop}, yields the final result:
\begin{align}\label{eq:timedepAmpFreq1}
G^\text{intra}_{11 \leftarrow 00} &= i \cC\cV^{-1}\sum_i\frac{g_i^2\Omega_i^2}{8\left(\omegaa^2-\Omega_i^2\right)^2}\left[\omegaa \tilde{f}_i\left(\mathbf{0},2\omegaa\right)+\int\frac{d\omega'}{2\pi}\frac{d^3k'}{(2\pi)^3} \frac{\omegaa\Omega_i^2 }{\left(\omegaa-\omega'\right)^2-\Omega_i^2}\tilde{f}_i\left(\bk',\omega'\right)\tilde{f}_i\left(-\bk',2\omegaa-\omega'\right)\right] \nonumber\\
& \quad\quad \quad  +  \cC^2\cV^{-2}\sum_{ij}\frac{g_i^2g_j^2\Omega_i^2\Omega_j^2}{64\left(\omegaa^2-\Omega_i^2\right)^2\left(\omegaa^2-\Omega_j^2\right)^2}\left[\omegaa^2 \tilde{f}_i\left(\mathbf{0},2\omegaa\right)\tilde{f}_j\left(\mathbf{0},0\right)\right] + \mathcal{O}(f^3),
\end{align}
where $\mathcal{V}$ is the volume of the medium. It is worth noting that in this process, where we consider two polaritons are emitted back-to-back, the medium modulation $f_i$ doesn't contribute with any additional momentum. Thus it is the homogeneous part of the modulation that is sampled. This is expected, as a pair of back-to-back polaritons automatically conserve momentum. Secondly, we are mostly interested in a periodically modulated medium, i.e. the dynamical Casimir effect, and therefore the zero frequency response is very small. Hence we can safely ignore the second line, which is proportional to $\tilde{f}_i(\mathbf{0},0)$. As for the vacuum radiation spectrum, we have two separate mechanisms here. One is a direct emission that depends only on the spectrum of the modulation $\tilde{f}$, this is the first term, whereas the second term explicitly depends on past events due to the integral over axillary frequency $\omega'$. This latter term allows for vacuum radiation resonances outside the spectrum of the modulation.

\subsubsection{Interbranch vacuum radiation}
For interbranch vacuum radiation we will first consider a slightly different driven amplitude, since in this case the two polaritons are distinguishable (at separate frequencies). Instead of Eq.~\eqref{eq:drivenAmp}, we must take the product of exciting one polariton into each of the branches. Thus we have 
\begin{align}\label{eq:drivenAmp1}
G^{J}_{11 \leftarrow 00} &= G^{J,\alpha}_{10\leftarrow 00}G^{J,\alpha'}_{01\leftarrow 00} \\ 
&= \left(\frac{1}{4\sqrt{\omegaa\omega_{\alpha'}}}\right) \int dt \; J^*_{\alpha}(t)e^{i\omegaa t}\int dt' \; J(t') e^{i\omega_{\alpha'} t'}\nonumber \\
&\quad\;\;\times\; \exp\left[-\frac{1}{4\omegaa}\int dt \int dt' J_\alpha(t)\cos\omegaa(t-t')J_\alpha^*(t')\right]\nonumber \\
&\quad\;\;\times\; \exp\left[-\frac{1}{4\omega_{\alpha'}}\int dt \int dt' J_{\alpha'}(t)\cos\omega_{\alpha'}(t-t')J_{\alpha'}^*(t')\right],
\end{align}
where we have added $\alpha$ or $\alpha'$ identifiers for clarity, and made sure that the process conserves momentum by involving a $\bk$ and a $-\bk$ polariton respectively. We can now substitute Eq.~\eqref{eq:drivenAmp1} into the perturbative procedure in Eq.~\eqref{eq:pertAmpCalc}, yielding
\begin{align}\label{eq:intra}
G^\text{inter}_{11 \leftarrow 00} = &\frac{i}{8\sqrt{\omegaa\omegaap}}\sigma^{\alpha\alpha'}_{\bk\bk}(\omegaa,\omegaap) - \frac{1}{128\sqrt{\omegaa^{3}\omegaap}}\sigma^{\alpha\alpha'}_{\bk\bk}(-\omegaa,\omegaap)\sigma^{\alpha\alpha}_{\bk\bk}(\omegaa,\omegaa),
\end{align}
where we have already neglected terms that would involve a factor of $\tilde{f}_i\left(\mathbf{0},0\right)$, for the same reason as for the intrabranch polaritons. Finally, we find the probability amplitude
\begin{align}\label{eq:intra1}
G^\text{inter}_{11 \leftarrow 00} &= i \sqrt{\cC\cCp}\cV^{-1}\sum_{i}\frac{g_i^2\Omega_i^2}{8\left(\omegaa^2-\Omega_i^2\right)\left(\omegaap^2-\Omega_i^2\right)}\bigg[\sqrt{\omegaa\omegaap} \tilde{f}_i\left(\mathbf{0},\omegaa+\omegaap\right) \nonumber\\
&\quad\quad\quad + \;\int\frac{d\omega'}{2\pi}\frac{d^3k'}{(2\pi)^3} \frac{\sqrt{\omegaa\omegaap}\Omega_i^2 }{\left(\omegaa-\omega'\right)^2-\Omega_i^2}\tilde{f}_i\left(\bk',\omega'\right)\tilde{f}_i\left(-\bk',\omegaa+\omegaap-\omega'\right)\bigg] \nonumber\\
& \quad\quad \quad  +  \cC\cCp\cV^{-2}\sum_{ij}\frac{g_i^2g_j^2\Omega_i^2\Omega_j^2\sqrt{\omegaa^3\omegaap} }{128\left(\omegaa^2-\Omega_i^2\right)\left(\omegaap^2-\Omega_i^2\right){\left(\omegaa^2-\Omega_j\right)^2}}\left[\tilde{f}_i\left(\mathbf{0},\omegaap-\omegaa\right)\tilde{f}_j\left(\mathbf{0},2\omegaa\right)\right] \nonumber \\
&\quad\quad\quad + \;\mathcal{O}(f^3).
\end{align}
We will return to this amplitude shortly, but it is once again worth noting that these interbranch processes open up the possibility for a variety of frequency mixing processes, as generally $\omega _\alpha$ and $\omega_{\alpha '}$ are at different frequencies. The spectrum of vacuum radiation depends directly on the spectrum of the modulation $f_i$, but due to the integral over $\omega'$ in the second line, also frequencies outside is possible. 

\subsubsection{Correlators}
As a quick aside, it is worth mentioning that correlators can be calculated with relative ease. This is done by applying the appropriate number of additional functional derivatives with respect to $J$ to the transition amplitude, before setting $J=0$. For instance, we can calculate the field-field correlator related to transitioning from vacuum to two back-to-back polaritons by
\begin{align*}
\langle 1_\bk, 1^*_\bk | A_{\bk\alpha}(t)&A^*_{\bq\alpha'}(\tau) |\Psi\rangle = \frac{\delta}{i\delta J^*_{\bk\alpha}(t)}\frac{\delta}{i\delta J_{\bq\alpha'}(\tau)}\left(e^{-iS_1\left[\frac{\delta}{i\delta J_{\bk\alpha}^*},\frac{\delta}{i\delta J_{\bk\alpha'}}\right]}G^J_{mn\leftarrow pq}(t_f,t_i)\right)_{J=0},
\end{align*}
where $|\Psi\rangle$ is the ground state $|0\rangle$ propagated with the time-modulated kernel, and $S_1$ is the action seen in the exponential when calculating the perturbative transition amplitudes in Eq.~\eqref{eq:pertAmpCalc}. We should note that it would here make sense not to consider a transition from the vacuum at $t_i = -\infty$ to an excited state $t_f = \infty$, but rather from $t_i~=~-T/2$ to $t_f~=~T/2$, and track the evolution of correlations as $T~=~t-\tau$ increases. However for the sake of brevity, we will not further discuss correlators in this work. 

\section{Frequency mixing of vacuum radiation}\label{sec:dispMix}
\begin{figure*}
\includegraphics[width=0.9\textwidth]{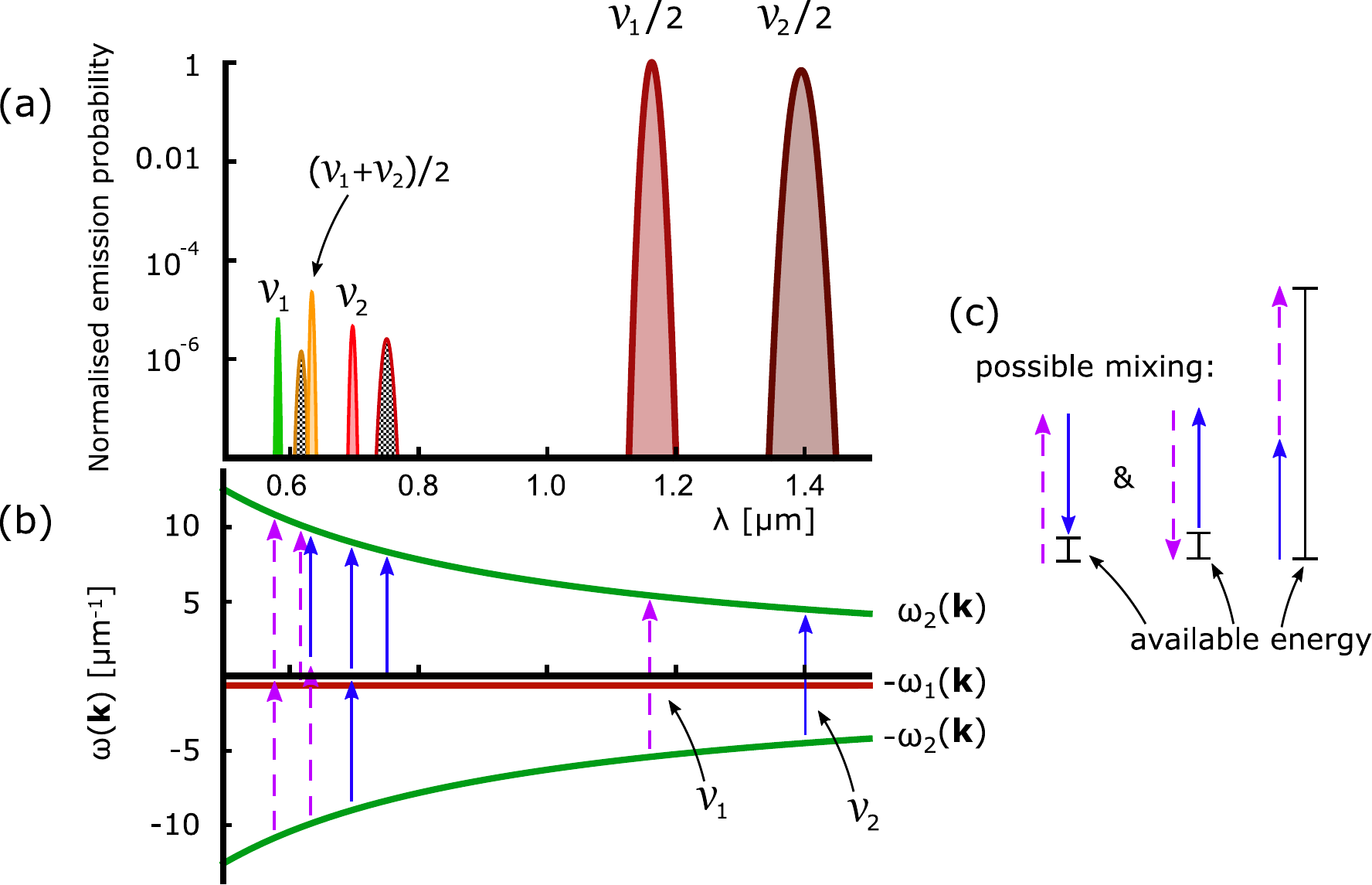}
\caption{\textbf{(a)} Excitation probability of two back-to-back polaritons in time-modulated fused silica, normalised to the maximum probability $4.4\times 10^{-6}$, as a function of \textit{vacuum} wavelength. Solid shading represents the intrabranch processes, whereas chequered shading denotes the interbranch transitions. Here we chose {$\delta n \simeq 10^{-3}$} by modulating the $\Omega_2$-resonance (near-visible-ultaviolet) at frequencies $\nu_1 = \Omega_2/5$ and $\nu_2 = \Omega_2/6$ for $\tau \simeq 42$ fs ($100$ fs full width at half maxima). Each resonance is labelled by the process, and we have ignored resonances that are outside the optical/infrared window. \textbf{(b)} The polariton branch of interest as a function of vacuum wavelength. An excitation process always involves a polariton-antipolariton pair, the latter having negative frequency. The time-modulation then provides the energy connecting the two branches (denoted by coloured arrows). \textbf{(c)} Illustration of the possible mixing processes at second order in perturbation theory.}\label{fig:probability}
\end{figure*}

In this section, we will explore further the dispersion-induced mixing processes mentioned briefly in the end of the last section. Specifically, let us consider a two-frequency time-dependence
\begin{align*}
f_i(\bx,t) = \epsilon_i \left(\cos\nu_1t + \cos\nu_2t\right)e^{-t^2/2\tau^2}
\end{align*} 
with the Fourier transform 
\begin{align*}
\tilde{f}_i(\bk,\omega) = \epsilon_i\tau\mathcal{V}\sqrt{\frac{\pi}{2}}&\bigg[e^{-\frac{1}{2}\tau^2\left(\omega-\nu_1\right)^2}+e^{-\frac{1}{2}\tau^2\left(\omega+\nu_1\right)^2} + e^{-\frac{1}{2}\tau^2\left(\omega-\nu_2\right)^2}+e^{-\frac{1}{2}\tau^2\left(\omega+\nu_2\right)^2}\bigg].
\end{align*}
In both the intrabranch [Eq.~\eqref{eq:timedepAmpFreq1}] and interbranch [Eq.~\eqref{eq:intra1}] sectors, there is an integral over an auxiliary frequency $\omega'$. We can evaluate this mixing integral in the large $\tau$-limit (i.e. modulating for many periods), yielding
\begin{align}\label{eq:nonlocInt}
I_\text{mixing}^{\alpha\alpha'} &= \int\frac{d\omega'}{2\pi}\frac{d^3k'}{(2\pi)^3} \frac{\Omega_i^2 }{\left(\omegaa-\omega'\right)^2-\Omega_i^2} \sqrt{\omegaa\omegaap}\tilde{f}_i\left(\bk',\omega'\right)\tilde{f}_i\left(-\bk',\omegaa+\omegaap-\omega'\right) \\
&\simeq \frac{\pi}{2}\epsilon_i^2 \mathcal{V} \left(\sqrt{\omegaa\omegaap}\tau\right) \sum_{ab} \frac{\Omega_i^2}{\left(\omegaa-\nu_a\right)^2-\Omega_i^2} e^{-\frac{1}{2}\tau^2\left(\omegaa+\omegaap-\nu_a-\nu_b\right)^2},
\end{align}
where we approximated $d\omega'/(2\pi) \simeq 1/\tau$, and where $a,b$ both runs over the possible frequencies $\{\pm\nu_1,\pm\nu_2\}$. This integral becomes significant when $\omegaa+\omegaap =  \nu_a + \nu_b$, leading to a plethora of mixed frequency resonances. 

For simplicity, we will here modulate the $m^\text{th}$-resonance of the medium ($\epsilon_i~=~\epsilon\delta_{im}$) for some large time $\tau\gg 1/\omegaa$. Now, recall Eq.~\eqref{eq:deltan}, where we can relate the size of the modulation $\epsilon_i$ to the change in the refractive index $\delta n${. It is however more convenient to work with changes to the permittivity $\varepsilon = n^2$, which we will denote $\delta\varepsilon$: the two are related through $\delta\varepsilon(\omega) = - 2 n(\omega)\delta n(\omega)$, and thus
\begin{align}
\delta \varepsilon(\omega) = \sum_i \frac{g_i^2\Omega_i^2}{\left(\omega^2-\Omega_i^2\right)^2}\epsilon_i = \frac{g_m^2\Omega_m^2}{\left(\omega^2-\Omega_m^2\right)^2}\epsilon.
\end{align} }
For this type of modulation, we see that the intrabranch amplitude can be re-written as
\begin{align}\label{eq:vacuumAmplitudeIntra}
G^{\text{intra}}_{11 \leftarrow 00} = i\sqrt{\frac{\pi}{128}} \cC &{\delta\varepsilon_{\omegaa}}\omegaa\tau\sum_{a}\bigg[e^{-\frac{\tau^2}{2}\left(2\omegaa-\nu_a\right)^2} + \sqrt{\frac{\pi}{2}}\sum_{b} {\delta\varepsilon_{\omegaa}} \frac{\left(\omegaa^2-\Omega_m^2\right)^2}{g_m^2\left(\left(\omegaa-\nu_a\right)^2-\Omega_m^2\right)} e^{-\frac{\tau^2}{2}\left(2\omegaa-\nu_a-\nu_b\right)^2}\bigg],
\end{align}
where we denote {$\delta\varepsilon_{\omegaa}= \delta \varepsilon(\omegaa)$}, and where we as before have truncated the expansion to $\mathcal{O}\left(\epsilon^3\right)$. Similarly, the interbranch transition amplitude is given by 
\begin{align}\label{eq:vacuumAmplitudeInter}
G^{\text{inter}}_{11 \leftarrow 00} = i\sqrt{\frac{\pi}{128}} \sqrt{\cC\cCp} &\sqrt{{\delta\varepsilon_{\omegaa}}{\delta\varepsilon_{\omegaap}}} \sqrt{\omegaa\omegaap}\tau\sum_{a}\bigg[e^{-\frac{\tau^2}{2}\left(\omegaa+\omegaap-\nu_a\right)^2} \nonumber \\
&+ \quad \sqrt{\frac{\pi}{2}}\sum_{b} \sqrt{{\delta\varepsilon_{\omegaa}}{\delta\varepsilon_{\omegaap}}} \frac{\left(\omegaa^2-\Omega_m^2\right)\left(\omegaap^2-\Omega_m^2\right)}{g_m^2\left(\left(\omegaa-\nu_a\right)^2-\Omega_m^2\right)} e^{-\frac{\tau^2}{2}\left(\omegaa+\omegaap-\nu_a-\nu_b\right)^2} \nonumber \\
&- \quad i\sqrt{\frac{\pi}{512}}\sum_{b} \cC{\delta\varepsilon_{\omegaa}}\omegaa\tau e^{-\frac{\tau^2}{2}\left(\omegaap-\omegaa-\nu_a\right)^2} e^{-\frac{\tau^2}{2}\left(2\omegaa-\nu_b\right)^2}\bigg].
\end{align}
It is worth noting that the interbranch resonances are suppressed in general, as they require both branches to be photon-like simultaneously (so that $\cC\cCp \not\simeq 0$). Consequently, the last two term of Eq.~\eqref{eq:vacuumAmplitudeInter} can safely be neglected, as they furthermore contribute at the next order in perturbation theory. These additional vacuum radiation resonances are nonetheless possible. We can now calculate the total excitation probability by $\left|G^{\text{intra}}_{11 \leftarrow 00}+G^{\text{inter}}_{11 \leftarrow 00}\right|^2$ \footnote{{Note that the total probability density for emission is given by 
\begin{align*}
\frac{dP}{d\cV} = \int \frac{d^3k}{(2\pi)^3}\left|G^{\text{intra}}_{11 \leftarrow 00}+G^{\text{inter}}_{11 \leftarrow 00}\right|^2.
\end{align*}
Also, there is no need for renormalising this integral, as we are considering differences between the occupation in each state, not the total occupation number in each.}}.

Let us at this point specify the medium as fused silica (as in Fig.~\ref{fig:branches}), and as we are usually interested in optical frequencies, we will modulate the first ultraviolet ($\Omega_2$) medium resonance only. Specifically, we let $\nu_1 = \Omega_2/5$ and $\nu_2 = \Omega_2/6$, and choose $\epsilon$ such that {$\delta n \simeq 10^{-3}$} (small but standard for fused silica). The associated probability spectrum can be seen in Fig.~\ref{fig:probability}(a), where solid and chequered shading denotes an intra- and interbranch processes respectively. The polariton branches of interest are shown in Fig.~\ref{fig:probability}(b), along with the relevant modulation terms.

As can be seen, the temporal modulation provides the energy to resonantly connect a polariton branch with some antipolariton branch, which causes polaritons to be emitted from the vacuum state. Only the $\omega_1$ and $\omega_2$ branches are at a comparable scale to the modulation frequency ($\propto\Omega_2$), and are thus the only ones into which vacuum radiation is emitted. There are nonetheless several different possibilities, where the modulation energy will match either $2\omega_2$ (intrabranch) or $\omega_2 + \omega_1$ (interbranch). This opens up for the possibility of frequency-mixed vacuum radiation, where the frequency of emitted vacuum radiation is given by a combination of the frequencies present in the system. Note however, that both polaritons in any given pair will however oscillate at the same frequency when measured outside the optical medium, as they are at the same wavelength, leading to a measured spectrum such as the one seen in Fig.~\ref{fig:probability}. 

Starting with the intrabranch resonances, we see the two expected dynamical Casimir-like resonances, that is, $\omega_2 = \nu_{1,2}/2$. However, we also see resonances at $\omega_2 = \nu_{1,2}$ as well as at the mixed frequency $\omega_2 = (\nu_1+\nu_2)/2$. The difference frequency would become relevant when $\omega_2 = \left|\nu_1-\nu_2\right|/2$, which is in the far infrared and is ignored here. Furthermore, the interbranch resonances contribute also, when $\omega_2 + \omega_1 = \nu_1$ and $\omega_2 + \omega_1 = \nu_2$, denoted as chequered shading with yellow and red solid line, respectively, in Fig.~\ref{fig:probability}.

This is reminiscent of nonlinear processes, where sum and difference frequency generation is commonplace \cite{boyd}. However, the system studied in this work is by assumption linear. In fact, these resonances in the spectrum of emitted vacuum radiation has much in common with the resonances of classical parametric oscillators. It is known that a stand-alone parametric oscillator with oscillator frequency $\Omega$ has a \textit{primary} resonance at $\Omega = \nu/2$ if driven at frequency $\nu$, and several \textit{sub-harmonic} resonances at $\Omega = \nu$ and $3\nu/2$, and so on, where the strength of each resonance down the line is significantly weaker than the last \cite{arnol2013mathematical}. Also, coupled parametric oscillators has been shown to exhibit a variety of \textit{combination} (i.e. frequency-mixing) resonances \cite{paraComb1,paraComb2,paraComb3}, closely connected to the interbranch processes discussed here.

However, this does not explain resonances of the form $\omegaa + \omegaap = \left|\nu_1 \pm \nu_2\right|$. The type of frequency mixing is of a different nature than the `combination' parametric resonances. {Instead, the mixing relates to the parametric driving of the system by the beating pattern formed by the different components of the modulation (in the time domain), which oscillates at frequencies outside its spectrum.} An example of this is the `superoscillations' studied in Ref.~\cite{angusSuperoscillations}, but is in this case of much familiar origin: The two waves, $\cos\nu_1 t$ and $\cos\nu_2 t$, beat at $(\nu_1 + \nu_2)/2$ and $(\nu_1-\nu_2)/2$. The system is however unable to absorb the energy represented by this beating pattern directly. Rather it is a two-stage (virtual) process where one quanta of energy is absorbed by the first modulation wave ($\cos\nu_1 t$), which is stored, while the second modulation wave ($\cos\nu_2 t$) either adds or removes another quanta of energy from the system. Note, the removing of energy comes from the absorption of an anti-quanta of the modulation wave. The total energy is then emitted in the form of two polaritons. 

{It is worth pointing out that had we instead chosen to temporally modulate the light-matter coupling strengths $g_i$ instead of the oscillator frequencies $\Omega_i$, it is easy to see that we would not get modifications to the oscillator propagator seen in Eq.~\eqref{eq:oscprop}, the last line of which is responsible for the time-nonlocal integral in Eq.~\eqref{eq:nonlocInt}. Instead this would act similarly to a driving force. We would indeed also find quantum vacuum radiation in this scenario, and to $\mathcal{O}(\epsilon)$ look very similar, fulfilling the condition $\omegaa+\omegaap = r\nu_{1,2}$ for some integer $r$ \cite{referee1ref2,referee1ref3}. However, we expect the contribution from both frequencies of the drive ($|\nu_1\pm\nu_2|$) to disappear in this case (at least to the same order in $\epsilon$).}

{
\section{Discussion and conclusion}\label{sec:conclusion}

In conclusion, we have studied quantum vacuum radiation excited by temporal changes to the resonance frequency of an optical medium. In particular, we have examined how the dispersive response affects the spectrum of emitted photons. We studied this with bulk media in mind, and specified fused silica as an example. We found that the delayed temporal response of the medium, responsible for dispersion, introduces frequency mixing to the system. The spectrum of emitted photons then takes on a character reminiscent of nonlinear optics, where both sum and difference frequency emission is possible.}

Specifically, we showed that photons are emitted when the sum of two polariton {branch} frequencies match a combination of modulation frequencies. This we found led to several quantum vacuum radiation resonances, including $\left|\nu_1 \pm \nu_2\right|/2$ as well as the usual dynamical Casimir-like emission at $\nu_1/2$ and $\nu_2/2$, when modulating the medium at frequencies $\nu_1$ and $\nu_2$. We note that the system is by assumption linear, as to not confuse this with a nonlinear phenomenon. {We found instead that there are two separate, linear, mechanisms by which frequencies can mix, related either to the \textit{energy emission} process or the \textit{energy absorption} process, or a combination thereof.

The mixing of polariton branch frequencies is a consequence of the nature of coupled systems having multiple modes of oscillation, which in this case are the polariton branches. In the most simple case when the optical medium only has a single resonance frequency $\Omega$, the two modes oscillate at frequencies
\begin{align*}
\omega_{\pm} = \left|\sqrt{\frac{1}{2}\left((k^2+\Omega^2+g^2)\pm\sqrt{(k^2+\Omega^2+g^2)^2-4 k^2\Omega^2}\right)}\right|.
\end{align*} 
It follows that any excitation in the system, and hence emitted vacuum radiation, must consist of some combination of $\omega_+$-polaritons and $\omega_-$-polaritons. In the case of fused silica, there are further branches, whose algebraic form is considerably more complicated, but the physics is the same. 

On the other hand, the mixing of drive frequencies (i.e. the sum/difference frequency peaks) has a more subtle origin, and is connected to the time-delayed response of the medium to changes in its resonance frequencies. We find that when modulated at multiple frequencies, say $\nu_1$ and $\nu_2$ simultaneously, the medium can absorb energy from the beating pattern formed between the two waves. This process relies on a time-delayed response to changes in the resonance frequency, as the medium must first absorb one quanta of energy from one drive (say $\nu_1$), and at a later time absorb a (anti)quanta of energy ($-$)$\nu_2$ from the second drive. The total energy of $|\nu_1 \pm \nu_2|$ is then emitted in form of polaritons, and as such energy conservation requires that $\omegaa+\omegaap = \nu_1\pm\nu_2$. }

In order to study this, we used a microscopic phenomenological model for electromagnetism {in an optical medium with a generic Sellmeier dispersion relation}. We quantised this using a path integral formalism. No approximations were made with regards to the delayed response, and dispersion was therefore fully taken into account. Within this framework, we induced a time-dependent change to the refractive index $n$ by weakly perturbing the resonance frequencies of medium. {The model is however extendible to include also temporal changes to other parameters of the optical medium, such as the density and dipolar coupling strengths. It is worth noting that this model relates most readily to experiments in bulk media, such as in Refs.~\cite{enzD1,enzD2,modFibre}, rather than the typical cavity set-up where polariton physics is more commonly discussed \cite{referee1ref2,referee1ref3,referee1ref4,referee1ref5,referee1ref6}. 

The origin of the time-dependent resonance frequencies has not been mentioned explicitly in this work, but has been kept general. Nonetheless, the results are directly applicable to experiments in which the temporal changes to the resonance frequency originates from the quadratic Stark shift (as discussed in Ref.~\cite{unruh}), i.e. $\Omega_i \rightarrow \Omega_i + \gamma E_\tm{pump}^2$, for some strong electric field $E_\tm{pump}$. Whilst this mechanism does introduce an actual nonlinearity to the system, we want to highlight that {this nonlinearity affects the pump beam only, and the physics of the quantum vacuum discussed here is at all times linear}, especially since typical vacuum electric field fluctuations are exceedingly weak. {Indeed, this is the same line of reasoning as some recent discussion of the overlap between nonlinear optics and Casimir-Polder physics \cite{robs}.} Therefore, the framework is applicable to experiments with strong electric fields propagating in bulk, or structured, media, such as the fibre experiment in Ref.~\cite{modFibre}. {In the context of these bulk media with a strong pump pulse, we expect the mixed-frequency quantum vacuum radiation discussed in this work to be readily observable. Whilst the mixing is indeed a second order effect, the fact that it allows us to shift the frequency of the vacuum radiation to ranges with better detector efficiencies, such as the optical to infra-red regime \cite{detectors}, greatly improves the observability of quantum vacuum radiation. Suppose that the fused silica slab in Fig.~\ref{fig:probability} is a thin film of roughly $100 \mu\tm{m}$ thickness, and considerably larger than the pump laser spot size $A_\tm{spot}$ in the transverse direction. We can estimate the number of photon pairs emitted per unit angle $d\theta$ as 
\begin{align}
\frac{dP}{d\theta} &= \int_{0}^\infty dk \; k \left|G^{\text{intra}}_{11 \leftarrow 00}+G^{\text{inter}}_{11 \leftarrow 00}\right|^2 \nonumber\\
&\simeq A_\tm{spot} \left(\frac{2\pi}{\tau}\right)\left(\frac{2\pi}{\lambda_\tm{mix}}\right)\left|G^{\text{intra}}_{11 \leftarrow 00}+G^{\text{inter}}_{11 \leftarrow 00}\right|^2(\lambda_\tm{mix}) \simeq 3\times 10^{-6},
\end{align} 
where we have in the last step assumed $A_\tm{spot} = 250 \mu\tm{m}$ as well as approximated $dk \simeq 2\pi/\tau$ and $\lambda_\tm{mix} \simeq 0.65\mu\tm{m}$. This radiation would be emitted in the orthogonal direction to the pump beam (i.e. the transverse plane). This is the emission per pulse, so a repetition rate of $1 \tm{ MHz}$ yields roughly three photon pairs per second, which is measurable with current technology \cite{detectors}, given that they can be out-coupled from the medium (an experimental challenge but not impossible). Importantly, this frequency mixing is off-set from any other frequency of the system, and is therefore unlikely to be filtered away (a common problem for quantum vacuum radiation).

In addition, the dispersion also allows you to choose to work at frequencies where the physics is sensitive to small changes to the optical parameters, such as close to the point where the group velocity dispersion is close to zero (a common point of interest in fibre optics \cite{agrawal}). In fact, we would argue that this is indeed the mechanics of photon pair production in Ref.~\cite{modFibre}, albeit this requires further analysis that is outside the scope of this work.}

Another experiment that relies on this mechanism is described in Ref.~\cite{AngusNic}, where the refractive index of a thin-film epsilon-near-zero metamaterial is changed rapidly in time, building on experiments performed in Refs.~\cite{enzD1,enzD2}. In light of the present results, additional physics can be expected associated with the linear frequency mixing mechanisms. {This work suggests that the probability of emission for mixed frequency vacuum radiation to be $\propto (\delta n)^2$, where $\delta n$ is the absolute change of the refractive index. A back of the envelope calculation for the conditions in epsilon-near-zero materials (where $\delta n \simeq 0.9$) suggests near unity probability of emitting quantum vacuum radiation where $\sim 20 \%$ of the photons emitted would be frequency mixed.} Further study is required however, since {this is clearly not a perturbative change to the refractive index, and }absorption cannot always be neglected. This present work does nonetheless indicate that rich physics can be explored in the spectrum of emitted quantum vacuum radiation, especially in experiments with large changes to the refractive index.}

Finally, we note that we expect this {vacuum radiation} mixing phenomenon to be rather general, occurring in any temporally modulated system that has {delayed temporal responses}, and we note also that it is related to the parametric resonances of the system.

\newpage
\appendix
{
\section{Normal mode expansion in a fibre-like scenario}\label{app:normalmodes}}
Suppose that the oscillators has a weak space dependence in the transverse plane, i.e.  $\Omega^2_i(\bx)~\equiv~\Omega_i^2(\brho) = \Omega_0^2+\Delta r^2$. This describes a medium where the refractive index is a function of the transverse coordinates only $n(\omega,\brho)$, i.e. a fibre. As is typical for a fibre, let us assume that the index starts at a background value in the centre and then decrease by a small amount at higher radii. Furthermore, a good approximation is a parabolic profile if we choose this to be smooth. In the $z$-direction, we have the usual momentum modes $u_\bk(\bx) \propto u_\bn(\brho)e^{i k z}$, and the transverse dynamics is governed by 
\begin{equation}\label{eq:harFibre}
\left[-\nabla_\perp^2+2\alpha_k r^2\right]u_\bn(\brho)= 2E_k u_\bn(\brho),
\end{equation} 
where 
\begin{equation*}
\alpha_k = \frac{1}{2}\Delta \sum_i\frac{g^2\omega^2}{(\omega^2-\Omega_i^2)^2}
\end{equation*}
and 
\begin{equation*}
E_k = \frac{1}{2}\left[\omega^2\left(1-\sum_i\frac{g_i^2}{\omega^2-\Omega_i^2}\right)-k^2\right]
\end{equation*}
plays the role of harmonic frequency and energy respectively in this effective Schr\"odinger equation. In this, we expanded for small $\Delta$. Solutions of Eq.~\eqref{eq:harFibre} are the usual Hermite-Gaussian functions
\begin{align*}
u_{nm}(\brho) = &\left(\frac{\alpha_k}{2^{n+m}n!m!\pi}\right)^{1/2}e^{-\alpha_k \rho^2/2}  \nonumber \\
&\;\;\;\times H_n\left(\sqrt{\alpha_k}x\right)H_m\left(\sqrt{\alpha_k}y\right),
\end{align*}
where $\brho = (x,y)$, $H_l$ are the Hermite polynomials, and $n,m$ are integers. This also defines our dispersion relation, as solutions of Eq.~\eqref{eq:harFibre} have to follow
\begin{equation*}
E_k = \left(n+m+\frac{1}{2}\right)\alpha_k,
\end{equation*}
or written explicitly
\begin{align*}
D[k,m,n,\omega] \equiv &\left[-k^2+\omega^2\left(1-\sum_i\frac{g_i^2}{\omega^2-\Omega_i^2}\right)\right]- \left(n+m+\frac{1}{2}\right) \Delta\sum_i\frac{g^2\omega^2}{(\omega^2-\Omega_i^2)^2} = 0.
\end{align*}
Here we note that it is appropriate to treat $\Delta$ perturbatively also when solving for polariton branches.

\vspace{1cm}

\section{Calculation of transition amplitude}\label{app:amplitude}
Varying the classical action in Eq.~\eqref{eq:effAcSimp} yields 
\begin{equation*}
\ddot{A}(t)+\omegaa^2 A(t) = J(t)
\end{equation*}
with the boundary conditions $A(t_f) = A_f$ and $A(t_i) = A_i$, and similarly for $A^*$. The solution can be separated into homogeneous and inhomogeneous parts, $A_\tm{H}$ and $A_\tm{I}$ respectively, where the boundary conditions for the inhomogeneous part are $A_\tm{I}(t_f) = A_\tm{I}(t_i) = 0$. This leads to 
\begin{align}\label{eq:EoM}
A_\tm{H}(t) &= \frac{1}{\sin\omegaa T}\left[A_f\sin\omegaa(t-t_i)+A_i\sin\omegaa(t_f-t)\right] \nonumber \\
A_\tm{I}(t) &= -\bigg[\frac{\sin\omegaa(t_f-t)}{\omegaa\sin\omegaa T}\int_{t_i}^{t}dt' \; \sin\omegaa(t'-t_i)J(t') + \frac{\sin\omegaa(t-t_i)}{\omegaa\sin\omegaa T}\int_{t}^{t_f}dt' \; \sin\omegaa(t_f-t')J(t')\bigg].
\end{align}
By integrating by parts and using the classical equations of motion, it is easy to show that
\begin{align}\label{eq:classAct1}
S_\tm{cl}[J,J^*] = \frac{1}{2}&\left[A^*_\tm{H}\dot{A}_\tm{H}\right]_{t_i}^{t_f}+\frac{1}{2}\left[A^*_\tm{H}\dot{A}_\tm{I}\right]_{t_i}^{t_f}+\frac{1}{2}\int_{t_i}^{t_f}dt \;J^*(t)\left[A_\tm{H}(t)+A_\tm{I}(t)\right].
\end{align}
After substituting in the solutions in Eq.~\eqref{eq:EoM} into Eq.~\eqref{eq:classAct1}, we arrive (after some algebra) at 
\begin{widetext}
\begin{align}\label{eq:classAct}
S_\tm{cl}[J,J^*] =& \;\frac{\omegaa}{2\sin\omegaa T}\left[\left(|A_f|^2+|A_i|^2\right)\cos\omegaa T - \left(A^*_f A_i + c.c.\right)\right] \nonumber \\
&+ \frac{1}{2\sin\omegaa T}\left[A_f \int_{t_i}^{t_f}dt \;\sin\omegaa\left(t-t_i\right)J^*(t) + A_i\int_{t_i}^{t_f}dt \;\sin\omegaa\left(t_f-t\right)J^*(t) + c.c.\right] \nonumber \\
&-\frac{1}{2\omegaa\sin\omegaa T}\left[\int_{t_i}^{t_f}dt\int_{t_i}^{t_f}dt' \; J^*(t)\sin\omegaa\left(t_f-t\right)\sin\omegaa\left(t'-t_i\right)J(t') + c.c.\right],
\end{align}
\end{widetext}
where $c.c.$ denotes the complex conjugate.

Finally, by splitting the field into classical and quantum components, that is $A = A_\tm{cl} + \eta$, one can show that the transition amplitude is given by 
\begin{align*}
\brak{A_f,t_f}{A_i,t_i} &= e^{i S_\tm{cl}}\int_{\eta(t_i)=0}^{\eta(t_f)=0} \mathcal{D}\eta\mathcal{D}\eta^* \; \exp\left(i S_\tm{eff}[0,0]\right) \nonumber \\
&= \cF(T)e^{i S_\tm{cl}}.
\end{align*}
The normalisation factor remaining can now be computed by using the time-translational invariance of the problem, leading to 
\begin{equation*}
\cF(T) = \brak{0,T}{0,0} = \left(\frac{\omegaa}{4\pi i \sin\omegaa T}\right).
\end{equation*}
We should note that as $T\rightarrow 0$, this propagator reduces to the usual 
\begin{align*}
\brak{A_f,t_f}{A_i,t_i}_J = \delta(A_f-A_i).
\end{align*}
Also, if we were to at this point add up all normal modes $\bk$, then in most scenarios the propagator is formally infinite. This is commonplace for a field theory, and usually demands for renormalisation \cite{srednicki,rovelli}, but these infinities cancel when we consider transition amplitudes.

\section{Photon wavefunction(al)s}\label{app:wavefunctions}
We will here present the wavefunctionals for polariton states. They take the form of $mn$-states where $m$ and $n$ denotes the number of $-\bk$ and $\bk$ photons respectively, and the total number of photons are given by $m+n$. We can first introduce the complex Hermite polynomials $H_{mn}(x^*,x)$ given in Refs.~\cite{complexHermite1,complexHermite2,complexHermite3}
\begin{equation}\label{eq:hermite}
H_{mn}(x^*,x) = \sum_{k=0}^{\min(m,n)}(-1)^{k}k! {m\choose{k}}{n\choose{k}}\left(x^*\right)^{m-k}x^{n-k},
\end{equation} 
leading to a compact form of the wavefunctionals:
\begin{widetext}
\begin{equation*}
\Psi_{mn}(A,t) = \sqrt{\frac{\omega_{\alpha}}{2\pi}}\frac{e^{-\omega_{\alpha}\left|A\right|^2/2}}{\sqrt{m!n!}}H_{mn}(\sqrt{\omegaa}A^*,\sqrt{\omegaa}A)e^{-i(m+n)\omega_\alpha t},
\end{equation*}
Explicitly, the $0$-, $1$- and $2$-photon states are:
\begin{align*}
\Psi_{00}(A_{\bk\alpha},t) &= \sqrt{\frac{\omega_{\alpha}}{2\pi}}\exp\left(-\omega_{\alpha}\left|A_{\bk\alpha}\right|^2/2\right)e^{-i\omega_\alpha t} \\
\Psi_{01\lor 10}(A_{\bk\alpha},t) &= \sqrt{\frac{\omega_{\alpha}}{2\pi}}\bigg[\sqrt{\omegaa}A_{\bk\alpha} \lor \sqrt{\omegaa}A^*_{\bk\alpha} \bigg] \exp\left(-\omega_{\alpha}\left|A_{\bk\alpha}\right|^2/2\right)e^{-2i\omega_\alpha t}  \\
\Psi_{02\lor 20 \lor 11}(A_{\bk\alpha},t) &= \sqrt{\frac{\omega_{\alpha}}{2\pi}}\bigg[\frac{\omegaa}
{\sqrt{2}}\left(A_{\bk\alpha}\right)^2 \lor \frac{\omegaa}
{\sqrt{2}}\left(A^*_{\bk\alpha}\right)^2 \lor \left(\omegaa|A_{\bk\alpha}|^2-1\right) \bigg]\exp\left(-\omega_{\alpha}\left|A_{\bk\alpha}\right|^2/2\right)e^{-3i\omega_\alpha t},
\end{align*}
\end{widetext}
where $\lor$ (i.e. `or') denotes the choice depending on what combination of $\bk$ and $-\bk$ modes are excited. Generalising Ref.~\cite{feynman}, this was derived by first calculating the transition amplitude
\begin{align*}
F(b,a) = &\int d^2A_{f}d^2A_{i}\;\phi_b^*(A_{f})\brak{A_f,t_f}{A_i,t_i}\big|_{J=0}\phi_a(A_{i}),
\end{align*}
between functionals of the form  
\begin{equation*}
\phi_a(A) = \sqrt{\frac{\omega_{\alpha}}{2\pi}}e^{-\omegaa\left|A-a\right|^2/2}.
\end{equation*}
This yields
\begin{equation*}
F(b,a) = \gamma e^{-\omegaa \left(|b|^2 + |a|^2\right)/4}\exp\left[{\frac{\gamma\omegaa}{4}(b a^* + b^* a)}\right],
\end{equation*}
where $\gamma = \exp(-i\omegaa T)$. By expanding this in orders of $\gamma$, we can derive the coefficients $\psi_{m,n}(a)$ given in the expansion 
\begin{equation*}
\phi_a(A) = \sum_{m,n} \psi_{m,n}(a)\Psi_{m,n}(A).
\end{equation*}
We thus find the coefficients $\psi_{m,n}(a)$ for expanding $\phi_a(A)$ in terms of the polariton Fock space state functionals $\Psi_{m,n}(A)$, given by
\begin{equation}\label{eq:appWave}
\psi_{m,n}(a) = e^{-\frac{\omegaa}{4}|a|^2}\left(\frac{\omega}{4}\right)^{\frac{m+n}{2}}\frac{\left(a^*\right)^m a^n}{\sqrt{m!n!}}.
\end{equation}
With this knowledge, we can now perform the above expansion and extract $\Psi_{m,n}(A)$. 

\section{Generating functional for transition amplitudes}\label{app:transition}
Here we will simply state the result of the integral in Eq.~\eqref{eq:genInt}:
\begin{align*}
F(b,a)_J = \gamma G^J_{00} &e^{-\frac{\omegaa}{4} \left(|b|^2 + |a|^2\right)}\exp\left[{\frac{\gamma\omegaa}{4}(b a^* + b^* a)}\right]\exp{\left[i\gamma\sqrt{\frac{\omegaa}{4}}\left(b\beta_+^* + b^* \beta_+\right)\right]}  \exp{\left[i\sqrt{\frac{\omegaa}{4}}\left(a\beta_-^* + a^* \beta_-\right)\right]}
\end{align*}
where $\gamma = \exp(-i\omegaa T)$, and 
\begin{equation*}
\beta_{\pm}^{(*)} = \frac{1}{\sqrt{4\omegaa}}\int_{t_i}^{t_f}dt \; e^{\pm i\omegaa (t-t_i)}J^{(*)}(t).
\end{equation*}

\begin{acknowledgments}
NW would like to acknowledge insightful discussions with Jo\~ao C. Pinto Barros, Hans Thor Hansson and Fabio Biancalana. NW acknowledges support from EPSRC CM-CDT Grant No. EP/L015110/1. P\"O acknowledges support from EPSRC grant No. EP/M024636/1. D.F. acknowledges financial support from the European Research Council under the European Unions Seventh Framework Programme Grant No. (FP/2007-17172013)/ERC GA 306559 and EPSRC (U.K., Grant No. EP/J00443X/1).
\end{acknowledgments}

\newpage
\bibliographystyle{apsrev4-1}

\begin{thebibliography}{90}%
\makeatletter
\providecommand \@ifxundefined [1]{%
 \@ifx{#1\undefined}
}%
\providecommand \@ifnum [1]{%
 \ifnum #1\expandafter \@firstoftwo
 \else \expandafter \@secondoftwo
 \fi
}%
\providecommand \@ifx [1]{%
 \ifx #1\expandafter \@firstoftwo
 \else \expandafter \@secondoftwo
 \fi
}%
\providecommand \natexlab [1]{#1}%
\providecommand \enquote  [1]{``#1''}%
\providecommand \bibnamefont  [1]{#1}%
\providecommand \bibfnamefont [1]{#1}%
\providecommand \citenamefont [1]{#1}%
\providecommand \href@noop [0]{\@secondoftwo}%
\providecommand \href [0]{\begingroup \@sanitize@url \@href}%
\providecommand \@href[1]{\@@startlink{#1}\@@href}%
\providecommand \@@href[1]{\endgroup#1\@@endlink}%
\providecommand \@sanitize@url [0]{\catcode `\\12\catcode `\$12\catcode
  `\&12\catcode `\#12\catcode `\^12\catcode `\_12\catcode `\%12\relax}%
\providecommand \@@startlink[1]{}%
\providecommand \@@endlink[0]{}%
\providecommand \url  [0]{\begingroup\@sanitize@url \@url }%
\providecommand \@url [1]{\endgroup\@href {#1}{\urlprefix }}%
\providecommand \urlprefix  [0]{URL }%
\providecommand \Eprint [0]{\href }%
\providecommand \doibase [0]{http://dx.doi.org/}%
\providecommand \selectlanguage [0]{\@gobble}%
\providecommand \bibinfo  [0]{\@secondoftwo}%
\providecommand \bibfield  [0]{\@secondoftwo}%
\providecommand \translation [1]{[#1]}%
\providecommand \BibitemOpen [0]{}%
\providecommand \bibitemStop [0]{}%
\providecommand \bibitemNoStop [0]{.\EOS\space}%
\providecommand \EOS [0]{\spacefactor3000\relax}%
\providecommand \BibitemShut  [1]{\csname bibitem#1\endcsname}%
\let\auto@bib@innerbib\@empty
\bibitem [{\citenamefont {Yablonovitch}(1989)}]{timerefrac1}%
  \BibitemOpen
  \bibfield  {author} {\bibinfo {author} {\bibfnamefont {E.}~\bibnamefont
  {Yablonovitch}},\ }\href {\doibase 10.1103/PhysRevLett.62.1742} {\bibfield
  {journal} {\bibinfo  {journal} {Phys. Rev. Lett.}\ }\textbf {\bibinfo
  {volume} {62}},\ \bibinfo {pages} {1742} (\bibinfo {year}
  {1989})}\BibitemShut {NoStop}%
\bibitem [{\citenamefont {Mendon\ifmmode~\mbox{\c{c}}\else \c{c}\fi{}a}\ \emph
  {et~al.}(2000)\citenamefont {Mendon\ifmmode~\mbox{\c{c}}\else \c{c}\fi{}a},
  \citenamefont {Guerreiro},\ and\ \citenamefont {Martins}}]{timerefrac3}%
  \BibitemOpen
  \bibfield  {author} {\bibinfo {author} {\bibfnamefont {J.~T.}\ \bibnamefont
  {Mendon\ifmmode~\mbox{\c{c}}\else \c{c}\fi{}a}}, \bibinfo {author}
  {\bibfnamefont {A.}~\bibnamefont {Guerreiro}}, \ and\ \bibinfo {author}
  {\bibfnamefont {A.~M.}\ \bibnamefont {Martins}},\ }\href {\doibase
  10.1103/PhysRevA.62.033805} {\bibfield  {journal} {\bibinfo  {journal} {Phys.
  Rev. A}\ }\textbf {\bibinfo {volume} {62}},\ \bibinfo {pages} {033805}
  (\bibinfo {year} {2000})}\BibitemShut {NoStop}%
\bibitem [{\citenamefont {Mendon{\c{c}}a}(2000)}]{timerefracBook}%
  \BibitemOpen
  \bibfield  {author} {\bibinfo {author} {\bibfnamefont {J.~T.}\ \bibnamefont
  {Mendon{\c{c}}a}},\ }\href@noop {} {\emph {\bibinfo {title} {Theory of photon
  acceleration}}}\ (\bibinfo  {publisher} {CRC Press},\ \bibinfo {year}
  {2000})\BibitemShut {NoStop}%
\bibitem [{\citenamefont {Mendon{\c{c}}a}\ and\ \citenamefont
  {Shukla}(2002)}]{timerefrac4}%
  \BibitemOpen
  \bibfield  {author} {\bibinfo {author} {\bibfnamefont {J.~T.}\ \bibnamefont
  {Mendon{\c{c}}a}}\ and\ \bibinfo {author} {\bibfnamefont {P.~K.}\
  \bibnamefont {Shukla}},\ }\href
  {http://stacks.iop.org/1402-4896/65/i=2/a=006} {\bibfield  {journal}
  {\bibinfo  {journal} {Physica Scripta}\ }\textbf {\bibinfo {volume} {65}},\
  \bibinfo {pages} {160} (\bibinfo {year} {2002})}\BibitemShut {NoStop}%
\bibitem [{\citenamefont {Mendon\ifmmode~\mbox{\c{c}}\else \c{c}\fi{}a}\ and\
  \citenamefont {Guerreiro}(2005)}]{timerefrac2}%
  \BibitemOpen
  \bibfield  {author} {\bibinfo {author} {\bibfnamefont {J.~T.}\ \bibnamefont
  {Mendon\ifmmode~\mbox{\c{c}}\else \c{c}\fi{}a}}\ and\ \bibinfo {author}
  {\bibfnamefont {A.}~\bibnamefont {Guerreiro}},\ }\href {\doibase
  10.1103/PhysRevA.72.063805} {\bibfield  {journal} {\bibinfo  {journal} {Phys.
  Rev. A}\ }\textbf {\bibinfo {volume} {72}},\ \bibinfo {pages} {063805}
  (\bibinfo {year} {2005})}\BibitemShut {NoStop}%
\bibitem [{\citenamefont {Liberati}\ \emph {et~al.}(2012)\citenamefont
  {Liberati}, \citenamefont {Prain},\ and\ \citenamefont
  {Visser}}]{AngusLiberati}%
  \BibitemOpen
  \bibfield  {author} {\bibinfo {author} {\bibfnamefont {S.}~\bibnamefont
  {Liberati}}, \bibinfo {author} {\bibfnamefont {A.}~\bibnamefont {Prain}}, \
  and\ \bibinfo {author} {\bibfnamefont {M.}~\bibnamefont {Visser}},\ }\href
  {\doibase 10.1103/PhysRevD.85.084014} {\bibfield  {journal} {\bibinfo
  {journal} {Phys. Rev. D}\ }\textbf {\bibinfo {volume} {85}},\ \bibinfo
  {pages} {084014} (\bibinfo {year} {2012})}\BibitemShut {NoStop}%
\bibitem [{\citenamefont {Prain}\ \emph {et~al.}(2017)\citenamefont {Prain},
  \citenamefont {Vezzoli}, \citenamefont {Westerberg}, \citenamefont {Roger},\
  and\ \citenamefont {Faccio}}]{AngusNic}%
  \BibitemOpen
  \bibfield  {author} {\bibinfo {author} {\bibfnamefont {A.}~\bibnamefont
  {Prain}}, \bibinfo {author} {\bibfnamefont {S.}~\bibnamefont {Vezzoli}},
  \bibinfo {author} {\bibfnamefont {N.}~\bibnamefont {Westerberg}}, \bibinfo
  {author} {\bibfnamefont {T.}~\bibnamefont {Roger}}, \ and\ \bibinfo {author}
  {\bibfnamefont {D.}~\bibnamefont {Faccio}},\ }\href {\doibase
  10.1103/PhysRevLett.118.133904} {\bibfield  {journal} {\bibinfo  {journal}
  {Phys. Rev. Lett.}\ }\textbf {\bibinfo {volume} {118}},\ \bibinfo {pages}
  {133904} (\bibinfo {year} {2017})}\BibitemShut {NoStop}%
\bibitem [{\citenamefont {Vezzoli}\ \emph
  {et~al.}(2018{\natexlab{a}})\citenamefont {Vezzoli}, \citenamefont {Bruno},
  \citenamefont {DeVault}, \citenamefont {Roger}, \citenamefont {Shalaev},
  \citenamefont {Boltasseva}, \citenamefont {Ferrera}, \citenamefont {Clerici},
  \citenamefont {Dubietis},\ and\ \citenamefont {Faccio}}]{shalaev}%
  \BibitemOpen
  \bibfield  {author} {\bibinfo {author} {\bibfnamefont {S.}~\bibnamefont
  {Vezzoli}}, \bibinfo {author} {\bibfnamefont {V.}~\bibnamefont {Bruno}},
  \bibinfo {author} {\bibfnamefont {C.}~\bibnamefont {DeVault}}, \bibinfo
  {author} {\bibfnamefont {T.}~\bibnamefont {Roger}}, \bibinfo {author}
  {\bibfnamefont {V.~M.}\ \bibnamefont {Shalaev}}, \bibinfo {author}
  {\bibfnamefont {A.}~\bibnamefont {Boltasseva}}, \bibinfo {author}
  {\bibfnamefont {M.}~\bibnamefont {Ferrera}}, \bibinfo {author} {\bibfnamefont
  {M.}~\bibnamefont {Clerici}}, \bibinfo {author} {\bibfnamefont
  {A.}~\bibnamefont {Dubietis}}, \ and\ \bibinfo {author} {\bibfnamefont
  {D.}~\bibnamefont {Faccio}},\ }\href {\doibase
  10.1103/PhysRevLett.120.043902} {\bibfield  {journal} {\bibinfo  {journal}
  {Phys. Rev. Lett.}\ }\textbf {\bibinfo {volume} {120}},\ \bibinfo {pages}
  {043902} (\bibinfo {year} {2018}{\natexlab{a}})}\BibitemShut {NoStop}%
\bibitem [{\citenamefont {Landau}\ \emph {et~al.}(2013)\citenamefont {Landau},
  \citenamefont {Bell}, \citenamefont {Kearsley}, \citenamefont {Pitaevskii},
  \citenamefont {Lifshitz},\ and\ \citenamefont {Sykes}}]{landau}%
  \BibitemOpen
  \bibfield  {author} {\bibinfo {author} {\bibfnamefont {L.~D.}\ \bibnamefont
  {Landau}}, \bibinfo {author} {\bibfnamefont {J.}~\bibnamefont {Bell}},
  \bibinfo {author} {\bibfnamefont {M.}~\bibnamefont {Kearsley}}, \bibinfo
  {author} {\bibfnamefont {L.}~\bibnamefont {Pitaevskii}}, \bibinfo {author}
  {\bibfnamefont {E.}~\bibnamefont {Lifshitz}}, \ and\ \bibinfo {author}
  {\bibfnamefont {J.}~\bibnamefont {Sykes}},\ }\href@noop {} {\emph {\bibinfo
  {title} {Electrodynamics of continuous media}}},\ Vol.~\bibinfo {volume} {8}\
  (\bibinfo  {publisher} {Elsevier},\ \bibinfo {year} {2013})\BibitemShut
  {NoStop}%
\bibitem [{\citenamefont {Jackson}(1999)}]{jackson}%
  \BibitemOpen
  \bibfield  {author} {\bibinfo {author} {\bibfnamefont {J.~D.}\ \bibnamefont
  {Jackson}},\ }\href@noop {} {\emph {\bibinfo {title} {Classical
  electrodynamics}}}\ (\bibinfo  {publisher} {Wiley},\ \bibinfo {year}
  {1999})\BibitemShut {NoStop}%
\bibitem [{\citenamefont {Knoll}\ \emph {et~al.}(2001)\citenamefont {Knoll},
  \citenamefont {Scheel},\ and\ \citenamefont {Welsch}}]{MacroQEDOverview1}%
  \BibitemOpen
  \bibfield  {author} {\bibinfo {author} {\bibfnamefont {L.}~\bibnamefont
  {Knoll}}, \bibinfo {author} {\bibfnamefont {S.}~\bibnamefont {Scheel}}, \
  and\ \bibinfo {author} {\bibfnamefont {D.-G.}\ \bibnamefont {Welsch}},\
  }\href@noop {} {\bibfield  {journal} {\bibinfo  {journal} {Coherence and
  Statistics of Photons and Atoms, J. Perina (ed.), Wiley-Interscience}\ }
  (\bibinfo {year} {2001})}\BibitemShut {NoStop}%
\bibitem [{\citenamefont {Scheel}\ and\ \citenamefont
  {Buhmann}(2008)}]{MacroQEDOverview2}%
  \BibitemOpen
  \bibfield  {author} {\bibinfo {author} {\bibfnamefont {S.}~\bibnamefont
  {Scheel}}\ and\ \bibinfo {author} {\bibfnamefont {S.}~\bibnamefont
  {Buhmann}},\ }\href {\doibase 10.2478/v10155-010-0092-x} {\bibfield
  {journal} {\bibinfo  {journal} {Acta Physica Slovaca}\ }\textbf {\bibinfo
  {volume} {58}},\ \bibinfo {pages} {675 } (\bibinfo {year}
  {2008})}\BibitemShut {NoStop}%
\bibitem [{\citenamefont {Hopfield}(1958)}]{hopfield}%
  \BibitemOpen
  \bibfield  {author} {\bibinfo {author} {\bibfnamefont {J.~J.}\ \bibnamefont
  {Hopfield}},\ }\href {\doibase 10.1103/PhysRev.112.1555} {\bibfield
  {journal} {\bibinfo  {journal} {Phys. Rev.}\ }\textbf {\bibinfo {volume}
  {112}},\ \bibinfo {pages} {1555} (\bibinfo {year} {1958})}\BibitemShut
  {NoStop}%
\bibitem [{\citenamefont {Huttner}\ and\ \citenamefont
  {Barnett}(1992)}]{barnett}%
  \BibitemOpen
  \bibfield  {author} {\bibinfo {author} {\bibfnamefont {B.}~\bibnamefont
  {Huttner}}\ and\ \bibinfo {author} {\bibfnamefont {S.~M.}\ \bibnamefont
  {Barnett}},\ }\href {\doibase 10.1103/PhysRevA.46.4306} {\bibfield  {journal}
  {\bibinfo  {journal} {Phys. Rev. A}\ }\textbf {\bibinfo {volume} {46}},\
  \bibinfo {pages} {4306} (\bibinfo {year} {1992})}\BibitemShut {NoStop}%
\bibitem [{\citenamefont {Philbin}(2010)}]{philbin}%
  \BibitemOpen
  \bibfield  {author} {\bibinfo {author} {\bibfnamefont {T.~G.}\ \bibnamefont
  {Philbin}},\ }\href {http://stacks.iop.org/1367-2630/12/i=12/a=123008}
  {\bibfield  {journal} {\bibinfo  {journal} {New Journal of Physics}\ }\textbf
  {\bibinfo {volume} {12}},\ \bibinfo {pages} {123008} (\bibinfo {year}
  {2010})}\BibitemShut {NoStop}%
\bibitem [{\citenamefont {Simpson}\ and\ \citenamefont
  {Leonhardt}(2015)}]{forcesOnVacuum}%
  \BibitemOpen
  \bibfield  {author} {\bibinfo {author} {\bibfnamefont {W.~M.}\ \bibnamefont
  {Simpson}}\ and\ \bibinfo {author} {\bibfnamefont {U.}~\bibnamefont
  {Leonhardt}},\ }\href@noop {} {\emph {\bibinfo {title} {Forces of the quantum
  vacuum: An Introduction to Casimir Physics}}}\ (\bibinfo  {publisher} {World
  Scientific Publishing Company},\ \bibinfo {year} {2015})\BibitemShut
  {NoStop}%
\bibitem [{\citenamefont {Arnol'd}(2013)}]{arnol2013mathematical}%
  \BibitemOpen
  \bibfield  {author} {\bibinfo {author} {\bibfnamefont {V.~I.}\ \bibnamefont
  {Arnol'd}},\ }\href@noop {} {\emph {\bibinfo {title} {Mathematical methods of
  classical mechanics}}},\ Vol.~\bibinfo {volume} {60}\ (\bibinfo  {publisher}
  {Springer Science \& Business Media},\ \bibinfo {year} {2013})\BibitemShut
  {NoStop}%
\bibitem [{\citenamefont {Hsu}(1963)}]{paraComb1}%
  \BibitemOpen
  \bibfield  {author} {\bibinfo {author} {\bibfnamefont {C.}~\bibnamefont
  {Hsu}},\ }\href {\doibase 10.1115/1.3636563} {\bibfield  {journal} {\bibinfo
  {journal} {Journal of Applied Mechanics}\ }\textbf {\bibinfo {volume} {30}},\
  \bibinfo {pages} {367} (\bibinfo {year} {1963})}\BibitemShut {NoStop}%
\bibitem [{\citenamefont {Nayfeh}\ and\ \citenamefont
  {Mook}(1977)}]{paraComb2}%
  \BibitemOpen
  \bibfield  {author} {\bibinfo {author} {\bibfnamefont {A.~H.}\ \bibnamefont
  {Nayfeh}}\ and\ \bibinfo {author} {\bibfnamefont {D.~T.}\ \bibnamefont
  {Mook}},\ }\href {\doibase 10.1121/1.381535} {\bibfield  {journal} {\bibinfo
  {journal} {The Journal of the Acoustical Society of America}\ }\textbf
  {\bibinfo {volume} {62}},\ \bibinfo {pages} {375} (\bibinfo {year}
  {1977})}\BibitemShut {NoStop}%
\bibitem [{\citenamefont {Baskaran}\ and\ \citenamefont
  {Turner}(2003)}]{paraMech1}%
  \BibitemOpen
  \bibfield  {author} {\bibinfo {author} {\bibfnamefont {R.}~\bibnamefont
  {Baskaran}}\ and\ \bibinfo {author} {\bibfnamefont {K.~L.}\ \bibnamefont
  {Turner}},\ }\href {\doibase 10.1088/0960-1317/13/5/323} {\bibfield
  {journal} {\bibinfo  {journal} {Journal of Micromechanics and
  Microengineering}\ }\textbf {\bibinfo {volume} {13}},\ \bibinfo {pages} {701}
  (\bibinfo {year} {2003})}\BibitemShut {NoStop}%
\bibitem [{\citenamefont {Ganesan}\ \emph {et~al.}(2017)\citenamefont
  {Ganesan}, \citenamefont {Do},\ and\ \citenamefont {Seshia}}]{paraMech2}%
  \BibitemOpen
  \bibfield  {author} {\bibinfo {author} {\bibfnamefont {A.}~\bibnamefont
  {Ganesan}}, \bibinfo {author} {\bibfnamefont {C.}~\bibnamefont {Do}}, \ and\
  \bibinfo {author} {\bibfnamefont {A.}~\bibnamefont {Seshia}},\ }\href
  {\doibase 10.1209/0295-5075/119/10002} {\bibfield  {journal} {\bibinfo
  {journal} {EPL (Europhysics Letters)}\ }\textbf {\bibinfo {volume} {119}},\
  \bibinfo {pages} {10002} (\bibinfo {year} {2017})}\BibitemShut {NoStop}%
\bibitem [{\citenamefont {Jin}\ and\ \citenamefont {Song}(2005)}]{paraFluid1}%
  \BibitemOpen
  \bibfield  {author} {\bibinfo {author} {\bibfnamefont {J.}~\bibnamefont
  {Jin}}\ and\ \bibinfo {author} {\bibfnamefont {Z.}~\bibnamefont {Song}},\
  }\href@noop {} {\bibfield  {journal} {\bibinfo  {journal} {Journal of Fluids
  and Structures}\ }\textbf {\bibinfo {volume} {20}},\ \bibinfo {pages} {763}
  (\bibinfo {year} {2005})}\BibitemShut {NoStop}%
\bibitem [{\citenamefont {Vioque}\ \emph {et~al.}(2010)\citenamefont {Vioque},
  \citenamefont {Champneys},\ and\ \citenamefont {Truman}}]{paraFluid2}%
  \BibitemOpen
  \bibfield  {author} {\bibinfo {author} {\bibfnamefont {J.~G.}\ \bibnamefont
  {Vioque}}, \bibinfo {author} {\bibfnamefont {A.~R.}\ \bibnamefont
  {Champneys}}, \ and\ \bibinfo {author} {\bibfnamefont {M.}~\bibnamefont
  {Truman}},\ }\href {\doibase 10.1007/BF03322555} {\bibfield  {journal}
  {\bibinfo  {journal} {SeMA Journal}\ }\textbf {\bibinfo {volume} {51}},\
  \bibinfo {pages} {63} (\bibinfo {year} {2010})}\BibitemShut {NoStop}%
\bibitem [{\citenamefont {Shukrinov}\ \emph {et~al.}(2016)\citenamefont
  {Shukrinov}, \citenamefont {Azemtsa-Donfack}, \citenamefont {Rahmonov},\ and\
  \citenamefont {Botha}}]{paraElec1}%
  \BibitemOpen
  \bibfield  {author} {\bibinfo {author} {\bibfnamefont {Y.~M.}\ \bibnamefont
  {Shukrinov}}, \bibinfo {author} {\bibfnamefont {H.}~\bibnamefont
  {Azemtsa-Donfack}}, \bibinfo {author} {\bibfnamefont {I.}~\bibnamefont
  {Rahmonov}}, \ and\ \bibinfo {author} {\bibfnamefont {A.}~\bibnamefont
  {Botha}},\ }\href {\doibase 10.1063/1.4954777} {\bibfield  {journal}
  {\bibinfo  {journal} {Low Temperature Physics}\ }\textbf {\bibinfo {volume}
  {42}},\ \bibinfo {pages} {446} (\bibinfo {year} {2016})}\BibitemShut
  {NoStop}%
\bibitem [{\citenamefont {Castellanos-Beltran}\ \emph
  {et~al.}(2008)\citenamefont {Castellanos-Beltran}, \citenamefont {Irwin},
  \citenamefont {Hilton}, \citenamefont {Vale},\ and\ \citenamefont
  {Lehnert}}]{paraElec2}%
  \BibitemOpen
  \bibfield  {author} {\bibinfo {author} {\bibfnamefont {M.}~\bibnamefont
  {Castellanos-Beltran}}, \bibinfo {author} {\bibfnamefont {K.}~\bibnamefont
  {Irwin}}, \bibinfo {author} {\bibfnamefont {G.}~\bibnamefont {Hilton}},
  \bibinfo {author} {\bibfnamefont {L.}~\bibnamefont {Vale}}, \ and\ \bibinfo
  {author} {\bibfnamefont {K.}~\bibnamefont {Lehnert}},\ }\href {\doibase
  10.1038/nphys1090} {\bibfield  {journal} {\bibinfo  {journal} {Nature
  Physics}\ }\textbf {\bibinfo {volume} {4}},\ \bibinfo {pages} {929} (\bibinfo
  {year} {2008})}\BibitemShut {NoStop}%
\bibitem [{\citenamefont {Chang}\ \emph {et~al.}(2017)\citenamefont {Chang},
  \citenamefont {Lomonosov}, \citenamefont {Janusonis}, \citenamefont {Vlasov},
  \citenamefont {Temnov},\ and\ \citenamefont {Tobey}}]{paraLM2}%
  \BibitemOpen
  \bibfield  {author} {\bibinfo {author} {\bibfnamefont {C.~L.}\ \bibnamefont
  {Chang}}, \bibinfo {author} {\bibfnamefont {A.~M.}\ \bibnamefont
  {Lomonosov}}, \bibinfo {author} {\bibfnamefont {J.}~\bibnamefont
  {Janusonis}}, \bibinfo {author} {\bibfnamefont {V.~S.}\ \bibnamefont
  {Vlasov}}, \bibinfo {author} {\bibfnamefont {V.~V.}\ \bibnamefont {Temnov}},
  \ and\ \bibinfo {author} {\bibfnamefont {R.~I.}\ \bibnamefont {Tobey}},\
  }\href {\doibase 10.1103/PhysRevB.95.060409} {\bibfield  {journal} {\bibinfo
  {journal} {Phys. Rev. B}\ }\textbf {\bibinfo {volume} {95}},\ \bibinfo
  {pages} {060409} (\bibinfo {year} {2017})}\BibitemShut {NoStop}%
\bibitem [{\citenamefont {Dodonov}\ and\ \citenamefont
  {Dodonov}(2008)}]{referee1ref1}%
  \BibitemOpen
  \bibfield  {author} {\bibinfo {author} {\bibfnamefont {V.}~\bibnamefont
  {Dodonov}}\ and\ \bibinfo {author} {\bibfnamefont {A.}~\bibnamefont
  {Dodonov}},\ }\href {\doibase https://doi.org/10.1088/1742-6596/99/1/012006}
  {\bibfield  {journal} {\bibinfo  {journal} {Journal of Physics: Conference
  Series}\ }\textbf {\bibinfo {volume} {99}},\ \bibinfo {pages} {012006}
  (\bibinfo {year} {2008})}\BibitemShut {NoStop}%
\bibitem [{\citenamefont {Brown}\ \emph {et~al.}(2019)\citenamefont {Brown},
  \citenamefont {Lowenstein},\ and\ \citenamefont {Mathur}}]{forcedVacuum}%
  \BibitemOpen
  \bibfield  {author} {\bibinfo {author} {\bibfnamefont {K.}~\bibnamefont
  {Brown}}, \bibinfo {author} {\bibfnamefont {A.}~\bibnamefont {Lowenstein}}, \
  and\ \bibinfo {author} {\bibfnamefont {H.}~\bibnamefont {Mathur}},\ }\href
  {\doibase 10.1103/PhysRevA.99.022504} {\bibfield  {journal} {\bibinfo
  {journal} {Phys. Rev. A}\ }\textbf {\bibinfo {volume} {99}},\ \bibinfo
  {pages} {022504} (\bibinfo {year} {2019})}\BibitemShut {NoStop}%
\bibitem [{\citenamefont {Mills}\ and\ \citenamefont
  {Burstein}(1974)}]{polaritonsOld}%
  \BibitemOpen
  \bibfield  {author} {\bibinfo {author} {\bibfnamefont {D.}~\bibnamefont
  {Mills}}\ and\ \bibinfo {author} {\bibfnamefont {E.}~\bibnamefont
  {Burstein}},\ }\href {\doibase 10.1088/0034-4885/37/7/001} {\bibfield
  {journal} {\bibinfo  {journal} {Reports on Progress in Physics}\ }\textbf
  {\bibinfo {volume} {37}},\ \bibinfo {pages} {817} (\bibinfo {year}
  {1974})}\BibitemShut {NoStop}%
\bibitem [{\citenamefont {Artoni}\ and\ \citenamefont
  {Birman}(1991)}]{quantumPolaritons}%
  \BibitemOpen
  \bibfield  {author} {\bibinfo {author} {\bibfnamefont {M.}~\bibnamefont
  {Artoni}}\ and\ \bibinfo {author} {\bibfnamefont {J.~L.}\ \bibnamefont
  {Birman}},\ }\href {\doibase 10.1103/PhysRevB.44.3736} {\bibfield  {journal}
  {\bibinfo  {journal} {Phys. Rev. B}\ }\textbf {\bibinfo {volume} {44}},\
  \bibinfo {pages} {3736} (\bibinfo {year} {1991})}\BibitemShut {NoStop}%
\bibitem [{\citenamefont {Deng}\ \emph {et~al.}(2010)\citenamefont {Deng},
  \citenamefont {Haug},\ and\ \citenamefont {Yamamoto}}]{polCond}%
  \BibitemOpen
  \bibfield  {author} {\bibinfo {author} {\bibfnamefont {H.}~\bibnamefont
  {Deng}}, \bibinfo {author} {\bibfnamefont {H.}~\bibnamefont {Haug}}, \ and\
  \bibinfo {author} {\bibfnamefont {Y.}~\bibnamefont {Yamamoto}},\ }\href
  {\doibase 10.1103/RevModPhys.82.1489} {\bibfield  {journal} {\bibinfo
  {journal} {Rev. Mod. Phys.}\ }\textbf {\bibinfo {volume} {82}},\ \bibinfo
  {pages} {1489} (\bibinfo {year} {2010})}\BibitemShut {NoStop}%
\bibitem [{\citenamefont {Carusotto}\ and\ \citenamefont
  {Ciuti}(2013)}]{fluidsOfLightCarusotto}%
  \BibitemOpen
  \bibfield  {author} {\bibinfo {author} {\bibfnamefont {I.}~\bibnamefont
  {Carusotto}}\ and\ \bibinfo {author} {\bibfnamefont {C.}~\bibnamefont
  {Ciuti}},\ }\href {\doibase 10.1103/RevModPhys.85.299} {\bibfield  {journal}
  {\bibinfo  {journal} {Rev. Mod. Phys.}\ }\textbf {\bibinfo {volume} {85}},\
  \bibinfo {pages} {299} (\bibinfo {year} {2013})}\BibitemShut {NoStop}%
\bibitem [{\citenamefont {Reiserer}\ and\ \citenamefont
  {Rempe}(2015)}]{cavityQED}%
  \BibitemOpen
  \bibfield  {author} {\bibinfo {author} {\bibfnamefont {A.}~\bibnamefont
  {Reiserer}}\ and\ \bibinfo {author} {\bibfnamefont {G.}~\bibnamefont
  {Rempe}},\ }\href {\doibase 10.1103/RevModPhys.87.1379} {\bibfield  {journal}
  {\bibinfo  {journal} {Rev. Mod. Phys.}\ }\textbf {\bibinfo {volume} {87}},\
  \bibinfo {pages} {1379} (\bibinfo {year} {2015})}\BibitemShut {NoStop}%
\bibitem [{\citenamefont {Portolan}\ \emph {et~al.}(2014)\citenamefont
  {Portolan}, \citenamefont {Einkemmer}, \citenamefont {V{\"o}r{\"o}s},
  \citenamefont {Weihs},\ and\ \citenamefont {Rabl}}]{cavityQED1}%
  \BibitemOpen
  \bibfield  {author} {\bibinfo {author} {\bibfnamefont {S.}~\bibnamefont
  {Portolan}}, \bibinfo {author} {\bibfnamefont {L.}~\bibnamefont {Einkemmer}},
  \bibinfo {author} {\bibfnamefont {Z.}~\bibnamefont {V{\"o}r{\"o}s}}, \bibinfo
  {author} {\bibfnamefont {G.}~\bibnamefont {Weihs}}, \ and\ \bibinfo {author}
  {\bibfnamefont {P.}~\bibnamefont {Rabl}},\ }\href {\doibase
  10.1088/1367-2630/16/6/063030} {\bibfield  {journal} {\bibinfo  {journal}
  {New Journal of Physics}\ }\textbf {\bibinfo {volume} {16}},\ \bibinfo
  {pages} {063030} (\bibinfo {year} {2014})}\BibitemShut {NoStop}%
\bibitem [{\citenamefont {Svidzinsky}\ \emph {et~al.}(2013)\citenamefont
  {Svidzinsky}, \citenamefont {Yuan},\ and\ \citenamefont {Scully}}]{paraLM1}%
  \BibitemOpen
  \bibfield  {author} {\bibinfo {author} {\bibfnamefont {A.~A.}\ \bibnamefont
  {Svidzinsky}}, \bibinfo {author} {\bibfnamefont {L.}~\bibnamefont {Yuan}}, \
  and\ \bibinfo {author} {\bibfnamefont {M.~O.}\ \bibnamefont {Scully}},\
  }\href {\doibase 10.1103/PhysRevX.3.041001} {\bibfield  {journal} {\bibinfo
  {journal} {Phys. Rev. X}\ }\textbf {\bibinfo {volume} {3}},\ \bibinfo {pages}
  {041001} (\bibinfo {year} {2013})}\BibitemShut {NoStop}%
\bibitem [{\citenamefont {Aspelmeyer}\ \emph {et~al.}(2014)\citenamefont
  {Aspelmeyer}, \citenamefont {Kippenberg},\ and\ \citenamefont
  {Marquardt}}]{optoMech}%
  \BibitemOpen
  \bibfield  {author} {\bibinfo {author} {\bibfnamefont {M.}~\bibnamefont
  {Aspelmeyer}}, \bibinfo {author} {\bibfnamefont {T.~J.}\ \bibnamefont
  {Kippenberg}}, \ and\ \bibinfo {author} {\bibfnamefont {F.}~\bibnamefont
  {Marquardt}},\ }\href {\doibase 10.1103/RevModPhys.86.1391} {\bibfield
  {journal} {\bibinfo  {journal} {Rev. Mod. Phys.}\ }\textbf {\bibinfo {volume}
  {86}},\ \bibinfo {pages} {1391} (\bibinfo {year} {2014})}\BibitemShut
  {NoStop}%
\bibitem [{\citenamefont {Lemonde}\ and\ \citenamefont
  {Clerk}(2015)}]{optoMech1}%
  \BibitemOpen
  \bibfield  {author} {\bibinfo {author} {\bibfnamefont {M.-A.}\ \bibnamefont
  {Lemonde}}\ and\ \bibinfo {author} {\bibfnamefont {A.~A.}\ \bibnamefont
  {Clerk}},\ }\href {\doibase 10.1103/PhysRevA.91.033836} {\bibfield  {journal}
  {\bibinfo  {journal} {Phys. Rev. A}\ }\textbf {\bibinfo {volume} {91}},\
  \bibinfo {pages} {033836} (\bibinfo {year} {2015})}\BibitemShut {NoStop}%
\bibitem [{\citenamefont {T{\"o}rm{\"a}}\ and\ \citenamefont
  {Barnes}(2014)}]{surfacePlasmonReview}%
  \BibitemOpen
  \bibfield  {author} {\bibinfo {author} {\bibfnamefont {P.}~\bibnamefont
  {T{\"o}rm{\"a}}}\ and\ \bibinfo {author} {\bibfnamefont {W.~L.}\ \bibnamefont
  {Barnes}},\ }\href {\doibase 10.1088/0034-4885/78/1/013901} {\bibfield
  {journal} {\bibinfo  {journal} {Reports on Progress in Physics}\ }\textbf
  {\bibinfo {volume} {78}},\ \bibinfo {pages} {013901} (\bibinfo {year}
  {2014})}\BibitemShut {NoStop}%
\bibitem [{\citenamefont {Silveri}\ \emph {et~al.}(2017)\citenamefont
  {Silveri}, \citenamefont {Tuorila}, \citenamefont {Thuneberg},\ and\
  \citenamefont {Paraoanu}}]{quantumModulationReview}%
  \BibitemOpen
  \bibfield  {author} {\bibinfo {author} {\bibfnamefont {M.}~\bibnamefont
  {Silveri}}, \bibinfo {author} {\bibfnamefont {J.}~\bibnamefont {Tuorila}},
  \bibinfo {author} {\bibfnamefont {E.}~\bibnamefont {Thuneberg}}, \ and\
  \bibinfo {author} {\bibfnamefont {G.}~\bibnamefont {Paraoanu}},\ }\href
  {\doibase 10.1088/1361-6633/aa5170} {\bibfield  {journal} {\bibinfo
  {journal} {Reports on Progress in Physics}\ }\textbf {\bibinfo {volume}
  {80}},\ \bibinfo {pages} {056002} (\bibinfo {year} {2017})}\BibitemShut
  {NoStop}%
\bibitem [{\citenamefont {Dalvit}\ \emph {et~al.}(2011)\citenamefont {Dalvit},
  \citenamefont {Neto},\ and\ \citenamefont {Mazzitelli}}]{DCEreview}%
  \BibitemOpen
  \bibfield  {author} {\bibinfo {author} {\bibfnamefont {D.~A.~R.}\
  \bibnamefont {Dalvit}}, \bibinfo {author} {\bibfnamefont {P.~A.~M.}\
  \bibnamefont {Neto}}, \ and\ \bibinfo {author} {\bibfnamefont {F.~D.}\
  \bibnamefont {Mazzitelli}},\ }\enquote {\bibinfo {title} {Fluctuations,
  dissipation and the dynamical casimir effect},}\ in\ \href {\doibase
  10.1007/978-3-642-20288-9_13} {\emph {\bibinfo {booktitle} {Casimir
  Physics}}},\ \bibinfo {editor} {edited by\ \bibinfo {editor} {\bibfnamefont
  {D.}~\bibnamefont {Dalvit}}, \bibinfo {editor} {\bibfnamefont
  {P.}~\bibnamefont {Milonni}}, \bibinfo {editor} {\bibfnamefont
  {D.}~\bibnamefont {Roberts}}, \ and\ \bibinfo {editor} {\bibfnamefont
  {F.}~\bibnamefont {da~Rosa}}}\ (\bibinfo  {publisher} {Springer Berlin
  Heidelberg},\ \bibinfo {address} {Berlin, Heidelberg},\ \bibinfo {year}
  {2011})\ pp.\ \bibinfo {pages} {419--457}\BibitemShut {NoStop}%
\bibitem [{\citenamefont {Nation}\ \emph {et~al.}(2012)\citenamefont {Nation},
  \citenamefont {Johansson}, \citenamefont {Blencowe},\ and\ \citenamefont
  {Nori}}]{circuitDCE}%
  \BibitemOpen
  \bibfield  {author} {\bibinfo {author} {\bibfnamefont {P.~D.}\ \bibnamefont
  {Nation}}, \bibinfo {author} {\bibfnamefont {J.~R.}\ \bibnamefont
  {Johansson}}, \bibinfo {author} {\bibfnamefont {M.~P.}\ \bibnamefont
  {Blencowe}}, \ and\ \bibinfo {author} {\bibfnamefont {F.}~\bibnamefont
  {Nori}},\ }\href {\doibase 10.1103/RevModPhys.84.1} {\bibfield  {journal}
  {\bibinfo  {journal} {Rev. Mod. Phys.}\ }\textbf {\bibinfo {volume} {84}},\
  \bibinfo {pages} {1} (\bibinfo {year} {2012})}\BibitemShut {NoStop}%
\bibitem [{\citenamefont {Wilson}\ \emph {et~al.}(2011)\citenamefont {Wilson},
  \citenamefont {Johansson}, \citenamefont {Pourkabirian}, \citenamefont
  {Simoen}, \citenamefont {Johansson}, \citenamefont {Duty}, \citenamefont
  {Nori},\ and\ \citenamefont {Delsing}}]{gjohansson}%
  \BibitemOpen
  \bibfield  {author} {\bibinfo {author} {\bibfnamefont {C.}~\bibnamefont
  {Wilson}}, \bibinfo {author} {\bibfnamefont {G.}~\bibnamefont {Johansson}},
  \bibinfo {author} {\bibfnamefont {A.}~\bibnamefont {Pourkabirian}}, \bibinfo
  {author} {\bibfnamefont {M.}~\bibnamefont {Simoen}}, \bibinfo {author}
  {\bibfnamefont {J.}~\bibnamefont {Johansson}}, \bibinfo {author}
  {\bibfnamefont {T.}~\bibnamefont {Duty}}, \bibinfo {author} {\bibfnamefont
  {F.}~\bibnamefont {Nori}}, \ and\ \bibinfo {author} {\bibfnamefont
  {P.}~\bibnamefont {Delsing}},\ }\href {\doibase 10.1038/nature10561}
  {\bibfield  {journal} {\bibinfo  {journal} {Nature}\ }\textbf {\bibinfo
  {volume} {479}},\ \bibinfo {pages} {376} (\bibinfo {year}
  {2011})}\BibitemShut {NoStop}%
\bibitem [{\citenamefont {Parker}(1969)}]{timeQFT1}%
  \BibitemOpen
  \bibfield  {author} {\bibinfo {author} {\bibfnamefont {L.}~\bibnamefont
  {Parker}},\ }\href {\doibase 10.1103/PhysRev.183.1057} {\bibfield  {journal}
  {\bibinfo  {journal} {Phys. Rev.}\ }\textbf {\bibinfo {volume} {183}},\
  \bibinfo {pages} {1057} (\bibinfo {year} {1969})}\BibitemShut {NoStop}%
\bibitem [{\citenamefont {Birrel}\ and\ \citenamefont
  {Davies}(1982)}]{birrelDavies}%
  \BibitemOpen
  \bibfield  {author} {\bibinfo {author} {\bibfnamefont {N.}~\bibnamefont
  {Birrel}}\ and\ \bibinfo {author} {\bibfnamefont {P.}~\bibnamefont
  {Davies}},\ }\href@noop {} {\emph {\bibinfo {title} {Quantum Field Theory in
  curved space-time}}}\ (\bibinfo  {publisher} {Cambridge University Press,
  Cambridge},\ \bibinfo {year} {1982})\BibitemShut {NoStop}%
\bibitem [{\citenamefont {Jacobson}(2005)}]{jacobson}%
  \BibitemOpen
  \bibfield  {author} {\bibinfo {author} {\bibfnamefont {T.}~\bibnamefont
  {Jacobson}},\ }in\ \href@noop {} {\emph {\bibinfo {booktitle} {Lectures on
  Quantum Gravity}}}\ (\bibinfo  {publisher} {Springer},\ \bibinfo {year}
  {2005})\ pp.\ \bibinfo {pages} {39--89}\BibitemShut {NoStop}%
\bibitem [{\citenamefont {Fedichev}\ and\ \citenamefont
  {Fischer}(2004)}]{uweCosmo}%
  \BibitemOpen
  \bibfield  {author} {\bibinfo {author} {\bibfnamefont {P.~O.}\ \bibnamefont
  {Fedichev}}\ and\ \bibinfo {author} {\bibfnamefont {U.~R.}\ \bibnamefont
  {Fischer}},\ }\href {\doibase 10.1103/PhysRevA.69.033602} {\bibfield
  {journal} {\bibinfo  {journal} {Phys. Rev. A}\ }\textbf {\bibinfo {volume}
  {69}},\ \bibinfo {pages} {033602} (\bibinfo {year} {2004})}\BibitemShut
  {NoStop}%
\bibitem [{\citenamefont {Barcel\'o}\ \emph {et~al.}(2003)\citenamefont
  {Barcel\'o}, \citenamefont {Liberati},\ and\ \citenamefont
  {Visser}}]{cosmoVisser}%
  \BibitemOpen
  \bibfield  {author} {\bibinfo {author} {\bibfnamefont {C.}~\bibnamefont
  {Barcel\'o}}, \bibinfo {author} {\bibfnamefont {S.}~\bibnamefont {Liberati}},
  \ and\ \bibinfo {author} {\bibfnamefont {M.}~\bibnamefont {Visser}},\ }\href
  {\doibase 10.1103/PhysRevA.68.053613} {\bibfield  {journal} {\bibinfo
  {journal} {Phys. Rev. A}\ }\textbf {\bibinfo {volume} {68}},\ \bibinfo
  {pages} {053613} (\bibinfo {year} {2003})}\BibitemShut {NoStop}%
\bibitem [{\citenamefont {Prain}\ \emph {et~al.}(2010)\citenamefont {Prain},
  \citenamefont {Fagnocchi},\ and\ \citenamefont {Liberati}}]{cosmoAngus}%
  \BibitemOpen
  \bibfield  {author} {\bibinfo {author} {\bibfnamefont {A.}~\bibnamefont
  {Prain}}, \bibinfo {author} {\bibfnamefont {S.}~\bibnamefont {Fagnocchi}}, \
  and\ \bibinfo {author} {\bibfnamefont {S.}~\bibnamefont {Liberati}},\ }\href
  {\doibase 10.1103/PhysRevD.82.105018} {\bibfield  {journal} {\bibinfo
  {journal} {Phys. Rev. D}\ }\textbf {\bibinfo {volume} {82}},\ \bibinfo
  {pages} {105018} (\bibinfo {year} {2010})}\BibitemShut {NoStop}%
\bibitem [{\citenamefont {Westerberg}\ \emph {et~al.}(2014)\citenamefont
  {Westerberg}, \citenamefont {Cacciatori}, \citenamefont {Belgiorno},
  \citenamefont {Dalla~Piazza},\ and\ \citenamefont {Faccio}}]{westerbergNJP}%
  \BibitemOpen
  \bibfield  {author} {\bibinfo {author} {\bibfnamefont {N.}~\bibnamefont
  {Westerberg}}, \bibinfo {author} {\bibfnamefont {S.}~\bibnamefont
  {Cacciatori}}, \bibinfo {author} {\bibfnamefont {F.}~\bibnamefont
  {Belgiorno}}, \bibinfo {author} {\bibfnamefont {F.}~\bibnamefont
  {Dalla~Piazza}}, \ and\ \bibinfo {author} {\bibfnamefont {D.}~\bibnamefont
  {Faccio}},\ }\href {http://stacks.iop.org/1367-2630/16/i=7/a=075003}
  {\bibfield  {journal} {\bibinfo  {journal} {New Journal of Physics}\ }\textbf
  {\bibinfo {volume} {16}},\ \bibinfo {pages} {075003} (\bibinfo {year}
  {2014})}\BibitemShut {NoStop}%
\bibitem [{\citenamefont {Ciuti}\ \emph {et~al.}(2005)\citenamefont {Ciuti},
  \citenamefont {Bastard},\ and\ \citenamefont {Carusotto}}]{referee1ref2}%
  \BibitemOpen
  \bibfield  {author} {\bibinfo {author} {\bibfnamefont {C.}~\bibnamefont
  {Ciuti}}, \bibinfo {author} {\bibfnamefont {G.}~\bibnamefont {Bastard}}, \
  and\ \bibinfo {author} {\bibfnamefont {I.}~\bibnamefont {Carusotto}},\ }\href
  {\doibase 10.1103/PhysRevB.72.115303} {\bibfield  {journal} {\bibinfo
  {journal} {Phys. Rev. B}\ }\textbf {\bibinfo {volume} {72}},\ \bibinfo
  {pages} {115303} (\bibinfo {year} {2005})}\BibitemShut {NoStop}%
\bibitem [{\citenamefont {Liberato}\ \emph {et~al.}(2007)\citenamefont
  {Liberato}, \citenamefont {Ciuti},\ and\ \citenamefont
  {Carusotto}}]{referee1ref3}%
  \BibitemOpen
  \bibfield  {author} {\bibinfo {author} {\bibfnamefont {S.~D.}\ \bibnamefont
  {Liberato}}, \bibinfo {author} {\bibfnamefont {C.}~\bibnamefont {Ciuti}}, \
  and\ \bibinfo {author} {\bibfnamefont {I.}~\bibnamefont {Carusotto}},\ }\href
  {\doibase 10.1103/PhysRevLett.98.103602} {\bibfield  {journal} {\bibinfo
  {journal} {Phys. Rev. Lett.}\ }\textbf {\bibinfo {volume} {98}},\ \bibinfo
  {pages} {103602} (\bibinfo {year} {2007})}\BibitemShut {NoStop}%
\bibitem [{\citenamefont {Auer}\ and\ \citenamefont
  {Burkard}(2012)}]{referee1ref4}%
  \BibitemOpen
  \bibfield  {author} {\bibinfo {author} {\bibfnamefont {A.}~\bibnamefont
  {Auer}}\ and\ \bibinfo {author} {\bibfnamefont {G.}~\bibnamefont {Burkard}},\
  }\href {\doibase 10.1103/PhysRevB.85.235140} {\bibfield  {journal} {\bibinfo
  {journal} {Phys. Rev. B}\ }\textbf {\bibinfo {volume} {85}},\ \bibinfo
  {pages} {235140} (\bibinfo {year} {2012})}\BibitemShut {NoStop}%
\bibitem [{\citenamefont {Stassi}\ \emph {et~al.}(2013)\citenamefont {Stassi},
  \citenamefont {Ridolfo}, \citenamefont {Di~Stefano}, \citenamefont
  {Hartmann},\ and\ \citenamefont {Savasta}}]{referee1ref5}%
  \BibitemOpen
  \bibfield  {author} {\bibinfo {author} {\bibfnamefont {R.}~\bibnamefont
  {Stassi}}, \bibinfo {author} {\bibfnamefont {A.}~\bibnamefont {Ridolfo}},
  \bibinfo {author} {\bibfnamefont {O.}~\bibnamefont {Di~Stefano}}, \bibinfo
  {author} {\bibfnamefont {M.~J.}\ \bibnamefont {Hartmann}}, \ and\ \bibinfo
  {author} {\bibfnamefont {S.}~\bibnamefont {Savasta}},\ }\href {\doibase
  10.1103/PhysRevLett.110.243601} {\bibfield  {journal} {\bibinfo  {journal}
  {Phys. Rev. Lett.}\ }\textbf {\bibinfo {volume} {110}},\ \bibinfo {pages}
  {243601} (\bibinfo {year} {2013})}\BibitemShut {NoStop}%
\bibitem [{\citenamefont {Dodonov}(2009)}]{referee1ref6}%
  \BibitemOpen
  \bibfield  {author} {\bibinfo {author} {\bibfnamefont {A.}~\bibnamefont
  {Dodonov}},\ }\bibfield  {booktitle} {\emph {\bibinfo {booktitle} {Journal of
  Physics: Conference Series}},\ }\href {\doibase
  https://doi.org/10.1088/1742-6596/161/1/012029} {\ \textbf {\bibinfo {volume}
  {161}},\ \bibinfo {pages} {012029} (\bibinfo {year} {2009})}\BibitemShut
  {NoStop}%
\bibitem [{\citenamefont {Kockum}\ \emph {et~al.}(2017)\citenamefont {Kockum},
  \citenamefont {Macr{\`\i}}, \citenamefont {Garziano}, \citenamefont
  {Savasta},\ and\ \citenamefont {Nori}}]{referee1ref7}%
  \BibitemOpen
  \bibfield  {author} {\bibinfo {author} {\bibfnamefont {A.~F.}\ \bibnamefont
  {Kockum}}, \bibinfo {author} {\bibfnamefont {V.}~\bibnamefont {Macr{\`\i}}},
  \bibinfo {author} {\bibfnamefont {L.}~\bibnamefont {Garziano}}, \bibinfo
  {author} {\bibfnamefont {S.}~\bibnamefont {Savasta}}, \ and\ \bibinfo
  {author} {\bibfnamefont {F.}~\bibnamefont {Nori}},\ }\href {\doibase
  https://doi.org/10.1038/s41598-017-04225-3} {\bibfield  {journal} {\bibinfo
  {journal} {Scientific reports}\ }\textbf {\bibinfo {volume} {7}},\ \bibinfo
  {pages} {5313} (\bibinfo {year} {2017})}\BibitemShut {NoStop}%
\bibitem [{\citenamefont {Barachati}\ \emph {et~al.}(2015)\citenamefont
  {Barachati}, \citenamefont {De~Liberato},\ and\ \citenamefont
  {K\'ena-Cohen}}]{excitonPolariton}%
  \BibitemOpen
  \bibfield  {author} {\bibinfo {author} {\bibfnamefont {F.}~\bibnamefont
  {Barachati}}, \bibinfo {author} {\bibfnamefont {S.}~\bibnamefont
  {De~Liberato}}, \ and\ \bibinfo {author} {\bibfnamefont {S.}~\bibnamefont
  {K\'ena-Cohen}},\ }\href {\doibase 10.1103/PhysRevA.92.033828} {\bibfield
  {journal} {\bibinfo  {journal} {Phys. Rev. A}\ }\textbf {\bibinfo {volume}
  {92}},\ \bibinfo {pages} {033828} (\bibinfo {year} {2015})}\BibitemShut
  {NoStop}%
\bibitem [{\citenamefont {Anappara}\ \emph {et~al.}(2009)\citenamefont
  {Anappara}, \citenamefont {De~Liberato}, \citenamefont {Tredicucci},
  \citenamefont {Ciuti}, \citenamefont {Biasiol}, \citenamefont {Sorba},\ and\
  \citenamefont {Beltram}}]{ultraStrongPolaritons}%
  \BibitemOpen
  \bibfield  {author} {\bibinfo {author} {\bibfnamefont {A.~A.}\ \bibnamefont
  {Anappara}}, \bibinfo {author} {\bibfnamefont {S.}~\bibnamefont
  {De~Liberato}}, \bibinfo {author} {\bibfnamefont {A.}~\bibnamefont
  {Tredicucci}}, \bibinfo {author} {\bibfnamefont {C.}~\bibnamefont {Ciuti}},
  \bibinfo {author} {\bibfnamefont {G.}~\bibnamefont {Biasiol}}, \bibinfo
  {author} {\bibfnamefont {L.}~\bibnamefont {Sorba}}, \ and\ \bibinfo {author}
  {\bibfnamefont {F.}~\bibnamefont {Beltram}},\ }\href {\doibase
  10.1103/PhysRevB.79.201303} {\bibfield  {journal} {\bibinfo  {journal} {Phys.
  Rev. B}\ }\textbf {\bibinfo {volume} {79}},\ \bibinfo {pages} {201303}
  (\bibinfo {year} {2009})}\BibitemShut {NoStop}%
\bibitem [{\citenamefont {de~Sousa}\ and\ \citenamefont
  {Dodonov}(2015)}]{microscopicCavityDynCasi}%
  \BibitemOpen
  \bibfield  {author} {\bibinfo {author} {\bibfnamefont {I.}~\bibnamefont
  {de~Sousa}}\ and\ \bibinfo {author} {\bibfnamefont {A.}~\bibnamefont
  {Dodonov}},\ }\href {\doibase 10.1088/1751-8113/48/24/245302} {\bibfield
  {journal} {\bibinfo  {journal} {Journal of Physics A: Mathematical and
  Theoretical}\ }\textbf {\bibinfo {volume} {48}},\ \bibinfo {pages} {245302}
  (\bibinfo {year} {2015})}\BibitemShut {NoStop}%
\bibitem [{\citenamefont {Belgiorno}\ \emph {et~al.}(2015)\citenamefont
  {Belgiorno}, \citenamefont {Cacciatori},\ and\ \citenamefont
  {Dalla~Piazza}}]{belgiorno}%
  \BibitemOpen
  \bibfield  {author} {\bibinfo {author} {\bibfnamefont {F.}~\bibnamefont
  {Belgiorno}}, \bibinfo {author} {\bibfnamefont {S.}~\bibnamefont
  {Cacciatori}}, \ and\ \bibinfo {author} {\bibfnamefont {F.}~\bibnamefont
  {Dalla~Piazza}},\ }\href@noop {} {\bibfield  {journal} {\bibinfo  {journal}
  {Physica Scripta}\ }\textbf {\bibinfo {volume} {91}},\ \bibinfo {pages}
  {015001} (\bibinfo {year} {2015})}\BibitemShut {NoStop}%
\bibitem [{\citenamefont {Naylor}(2015)}]{vacuumSurfacePlasmon}%
  \BibitemOpen
  \bibfield  {author} {\bibinfo {author} {\bibfnamefont {W.}~\bibnamefont
  {Naylor}},\ }\href {\doibase 10.1103/PhysRevA.91.053804} {\bibfield
  {journal} {\bibinfo  {journal} {Phys. Rev. A}\ }\textbf {\bibinfo {volume}
  {91}},\ \bibinfo {pages} {053804} (\bibinfo {year} {2015})}\BibitemShut
  {NoStop}%
\bibitem [{\citenamefont {Hizhnyakov}\ \emph {et~al.}(2016)\citenamefont
  {Hizhnyakov}, \citenamefont {Loot},\ and\ \citenamefont
  {Azizabadi}}]{DCEsurfacePlasmon}%
  \BibitemOpen
  \bibfield  {author} {\bibinfo {author} {\bibfnamefont {V.}~\bibnamefont
  {Hizhnyakov}}, \bibinfo {author} {\bibfnamefont {A.}~\bibnamefont {Loot}}, \
  and\ \bibinfo {author} {\bibfnamefont {S.}~\bibnamefont {Azizabadi}},\ }\href
  {\doibase 10.1007/s00339-016-9916-y} {\bibfield  {journal} {\bibinfo
  {journal} {Applied Physics A}\ }\textbf {\bibinfo {volume} {122}},\ \bibinfo
  {pages} {333} (\bibinfo {year} {2016})}\BibitemShut {NoStop}%
\bibitem [{\citenamefont {Caspani}\ \emph {et~al.}(2016)\citenamefont
  {Caspani}, \citenamefont {Kaipurath}, \citenamefont {Clerici}, \citenamefont
  {Ferrera}, \citenamefont {Roger}, \citenamefont {Kim}, \citenamefont
  {Kinsey}, \citenamefont {Pietrzyk}, \citenamefont {Di~Falco}, \citenamefont
  {Shalaev}, \citenamefont {Boltasseva},\ and\ \citenamefont {Faccio}}]{enzD2}%
  \BibitemOpen
  \bibfield  {author} {\bibinfo {author} {\bibfnamefont {L.}~\bibnamefont
  {Caspani}}, \bibinfo {author} {\bibfnamefont {R.~P.~M.}\ \bibnamefont
  {Kaipurath}}, \bibinfo {author} {\bibfnamefont {M.}~\bibnamefont {Clerici}},
  \bibinfo {author} {\bibfnamefont {M.}~\bibnamefont {Ferrera}}, \bibinfo
  {author} {\bibfnamefont {T.}~\bibnamefont {Roger}}, \bibinfo {author}
  {\bibfnamefont {J.}~\bibnamefont {Kim}}, \bibinfo {author} {\bibfnamefont
  {N.}~\bibnamefont {Kinsey}}, \bibinfo {author} {\bibfnamefont
  {M.}~\bibnamefont {Pietrzyk}}, \bibinfo {author} {\bibfnamefont
  {A.}~\bibnamefont {Di~Falco}}, \bibinfo {author} {\bibfnamefont {V.~M.}\
  \bibnamefont {Shalaev}}, \bibinfo {author} {\bibfnamefont {A.}~\bibnamefont
  {Boltasseva}}, \ and\ \bibinfo {author} {\bibfnamefont {D.}~\bibnamefont
  {Faccio}},\ }\href {\doibase 10.1103/PhysRevLett.116.233901} {\bibfield
  {journal} {\bibinfo  {journal} {Phys. Rev. Lett.}\ }\textbf {\bibinfo
  {volume} {116}},\ \bibinfo {pages} {233901} (\bibinfo {year}
  {2016})}\BibitemShut {NoStop}%
\bibitem [{\citenamefont {Vezzoli}\ \emph
  {et~al.}(2018{\natexlab{b}})\citenamefont {Vezzoli}, \citenamefont {Mussot},
  \citenamefont {Westerberg}, \citenamefont {Kudlinski}, \citenamefont {Saleh},
  \citenamefont {Prain}, \citenamefont {Biancalana}, \citenamefont {Lantz},\
  and\ \citenamefont {Faccio}}]{modFibre}%
  \BibitemOpen
  \bibfield  {author} {\bibinfo {author} {\bibfnamefont {S.}~\bibnamefont
  {Vezzoli}}, \bibinfo {author} {\bibfnamefont {A.}~\bibnamefont {Mussot}},
  \bibinfo {author} {\bibfnamefont {N.}~\bibnamefont {Westerberg}}, \bibinfo
  {author} {\bibfnamefont {A.}~\bibnamefont {Kudlinski}}, \bibinfo {author}
  {\bibfnamefont {H.~D.}\ \bibnamefont {Saleh}}, \bibinfo {author}
  {\bibfnamefont {A.}~\bibnamefont {Prain}}, \bibinfo {author} {\bibfnamefont
  {F.}~\bibnamefont {Biancalana}}, \bibinfo {author} {\bibfnamefont
  {E.}~\bibnamefont {Lantz}}, \ and\ \bibinfo {author} {\bibfnamefont
  {D.}~\bibnamefont {Faccio}},\ }\href {https://arxiv.org/abs/1811.04262}
  {\bibfield  {journal} {\bibinfo  {journal} {arXiv preprint arXiv:1811.04262}\
  } (\bibinfo {year} {2018}{\natexlab{b}})}\BibitemShut {NoStop}%
\bibitem [{\citenamefont {Kempf}\ and\ \citenamefont
  {Prain}(2017)}]{angusSuperoscillations}%
  \BibitemOpen
  \bibfield  {author} {\bibinfo {author} {\bibfnamefont {A.}~\bibnamefont
  {Kempf}}\ and\ \bibinfo {author} {\bibfnamefont {A.}~\bibnamefont {Prain}},\
  }\href {\doibase 10.1063/1.4996135} {\bibfield  {journal} {\bibinfo
  {journal} {Journal of Mathematical Physics}\ }\textbf {\bibinfo {volume}
  {58}},\ \bibinfo {pages} {082101} (\bibinfo {year} {2017})}\BibitemShut
  {NoStop}%
\bibitem [{\citenamefont {Engheta}(2013)}]{enz1}%
  \BibitemOpen
  \bibfield  {author} {\bibinfo {author} {\bibfnamefont {N.}~\bibnamefont
  {Engheta}},\ }\href {\doibase 10.1126/science.1235589} {\bibfield  {journal}
  {\bibinfo  {journal} {Science}\ }\textbf {\bibinfo {volume} {340}},\ \bibinfo
  {pages} {286} (\bibinfo {year} {2013})}\BibitemShut {NoStop}%
\bibitem [{\citenamefont {Liberal}\ and\ \citenamefont {Engheta}(2016)}]{enz2}%
  \BibitemOpen
  \bibfield  {author} {\bibinfo {author} {\bibfnamefont {I.}~\bibnamefont
  {Liberal}}\ and\ \bibinfo {author} {\bibfnamefont {N.}~\bibnamefont
  {Engheta}},\ }\href {\doibase 10.1364/OPN.27.7.000026} {\bibfield  {journal}
  {\bibinfo  {journal} {Opt. Photon. News}\ }\textbf {\bibinfo {volume} {27}},\
  \bibinfo {pages} {26} (\bibinfo {year} {2016})}\BibitemShut {NoStop}%
\bibitem [{\citenamefont {Liberal}\ and\ \citenamefont {Engheta}(2017)}]{enz3}%
  \BibitemOpen
  \bibfield  {author} {\bibinfo {author} {\bibfnamefont {I.}~\bibnamefont
  {Liberal}}\ and\ \bibinfo {author} {\bibfnamefont {N.}~\bibnamefont
  {Engheta}},\ }\href {\doibase 10.1073/pnas.1611924114} {\bibfield  {journal}
  {\bibinfo  {journal} {Proceedings of the National Academy of Sciences}\
  }\textbf {\bibinfo {volume} {114}},\ \bibinfo {pages} {822} (\bibinfo {year}
  {2017})}\BibitemShut {NoStop}%
\bibitem [{\citenamefont {Alam}\ \emph {et~al.}(2016)\citenamefont {Alam},
  \citenamefont {De~Leon},\ and\ \citenamefont {Boyd}}]{enzD1}%
  \BibitemOpen
  \bibfield  {author} {\bibinfo {author} {\bibfnamefont {M.~Z.}\ \bibnamefont
  {Alam}}, \bibinfo {author} {\bibfnamefont {I.}~\bibnamefont {De~Leon}}, \
  and\ \bibinfo {author} {\bibfnamefont {R.~W.}\ \bibnamefont {Boyd}},\ }\href
  {\doibase 10.1126/science.aae0330} {\bibfield  {journal} {\bibinfo  {journal}
  {Science}\ ,\ \bibinfo {pages} {aae0330}} (\bibinfo {year}
  {2016})}\BibitemShut {NoStop}%
\bibitem [{\citenamefont {Chiao}\ \emph {et~al.}(2004)\citenamefont {Chiao},
  \citenamefont {Hansson}, \citenamefont {Leinaas},\ and\ \citenamefont
  {Viefers}}]{hansson}%
  \BibitemOpen
  \bibfield  {author} {\bibinfo {author} {\bibfnamefont {R.~Y.}\ \bibnamefont
  {Chiao}}, \bibinfo {author} {\bibfnamefont {T.~H.}\ \bibnamefont {Hansson}},
  \bibinfo {author} {\bibfnamefont {J.~M.}\ \bibnamefont {Leinaas}}, \ and\
  \bibinfo {author} {\bibfnamefont {S.}~\bibnamefont {Viefers}},\ }\href
  {\doibase 10.1103/PhysRevA.69.063816} {\bibfield  {journal} {\bibinfo
  {journal} {Phys. Rev. A}\ }\textbf {\bibinfo {volume} {69}},\ \bibinfo
  {pages} {063816} (\bibinfo {year} {2004})}\BibitemShut {NoStop}%
\bibitem [{\citenamefont {Linder}\ \emph {et~al.}(2016)\citenamefont {Linder},
  \citenamefont {Sch\"utzhold},\ and\ \citenamefont {Unruh}}]{unruh}%
  \BibitemOpen
  \bibfield  {author} {\bibinfo {author} {\bibfnamefont {M.~F.}\ \bibnamefont
  {Linder}}, \bibinfo {author} {\bibfnamefont {R.}~\bibnamefont
  {Sch\"utzhold}}, \ and\ \bibinfo {author} {\bibfnamefont {W.~G.}\
  \bibnamefont {Unruh}},\ }\href {\doibase 10.1103/PhysRevD.93.104010}
  {\bibfield  {journal} {\bibinfo  {journal} {Phys. Rev. D}\ }\textbf {\bibinfo
  {volume} {93}},\ \bibinfo {pages} {104010} (\bibinfo {year}
  {2016})}\BibitemShut {NoStop}%
\bibitem [{\citenamefont {Boyd}(2003)}]{boyd}%
  \BibitemOpen
  \bibfield  {author} {\bibinfo {author} {\bibfnamefont {R.~W.}\ \bibnamefont
  {Boyd}},\ }\href@noop {} {\emph {\bibinfo {title} {Nonlinear Optics}}}\
  (\bibinfo  {publisher} {Academic Press},\ \bibinfo {year} {2003})\
  Chap.~\bibinfo {chapter} {1}\BibitemShut {NoStop}%
\bibitem [{\citenamefont {Hecht}(2002)}]{hecht}%
  \BibitemOpen
  \bibfield  {author} {\bibinfo {author} {\bibfnamefont {E.}~\bibnamefont
  {Hecht}},\ }\href@noop {} {\emph {\bibinfo {title} {Optics}}},\ \bibinfo
  {edition} {3rd}\ ed.\ (\bibinfo  {publisher} {Pearson},\ \bibinfo {year}
  {2002})\ Chap.~\bibinfo {chapter} {3}\BibitemShut {NoStop}%
\bibitem [{\citenamefont {Feynman}\ \emph {et~al.}(2010)\citenamefont
  {Feynman}, \citenamefont {Hibbs},\ and\ \citenamefont {Styer}}]{feynman}%
  \BibitemOpen
  \bibfield  {author} {\bibinfo {author} {\bibfnamefont {R.}~\bibnamefont
  {Feynman}}, \bibinfo {author} {\bibfnamefont {A.}~\bibnamefont {Hibbs}}, \
  and\ \bibinfo {author} {\bibfnamefont {D.}~\bibnamefont {Styer}},\
  }\href@noop {} {\emph {\bibinfo {title} {Quantum Mechanics and Path
  Integrals: Emended Edition}}}\ (\bibinfo  {publisher} {Dover Publications},\
  \bibinfo {year} {2010})\ Chap.~\bibinfo {chapter} {9}\BibitemShut {NoStop}%
\bibitem [{\citenamefont {Grosche}\ and\ \citenamefont
  {Steiner}(1998)}]{handbook}%
  \BibitemOpen
  \bibfield  {author} {\bibinfo {author} {\bibfnamefont {C.}~\bibnamefont
  {Grosche}}\ and\ \bibinfo {author} {\bibfnamefont {F.}~\bibnamefont
  {Steiner}},\ }\href@noop {} {\emph {\bibinfo {title} {Handbook of Feynman
  Path Integrals}}}\ (\bibinfo  {publisher} {Springer},\ \bibinfo {year}
  {1998})\ Chap.~\bibinfo {chapter} {6}\BibitemShut {NoStop}%
\bibitem [{\citenamefont {Calvo}\ \emph {et~al.}(2006)\citenamefont {Calvo},
  \citenamefont {Pic\'on},\ and\ \citenamefont {Bagan}}]{OAMQFT}%
  \BibitemOpen
  \bibfield  {author} {\bibinfo {author} {\bibfnamefont {G.~F.}\ \bibnamefont
  {Calvo}}, \bibinfo {author} {\bibfnamefont {A.}~\bibnamefont {Pic\'on}}, \
  and\ \bibinfo {author} {\bibfnamefont {E.}~\bibnamefont {Bagan}},\ }\href
  {\doibase 10.1103/PhysRevA.73.013805} {\bibfield  {journal} {\bibinfo
  {journal} {Phys. Rev. A}\ }\textbf {\bibinfo {volume} {73}},\ \bibinfo
  {pages} {013805} (\bibinfo {year} {2006})}\BibitemShut {NoStop}%
\bibitem [{\citenamefont {Goodman}(2005)}]{GoodmanOptics}%
  \BibitemOpen
  \bibfield  {author} {\bibinfo {author} {\bibfnamefont {J.~W.}\ \bibnamefont
  {Goodman}},\ }\href@noop {} {\emph {\bibinfo {title} {Introduction to Fourier
  optics}}}\ (\bibinfo  {publisher} {Roberts and Company Publishers},\ \bibinfo
  {year} {2005})\BibitemShut {NoStop}%
\bibitem [{\citenamefont {Loudon}(2000)}]{loudon}%
  \BibitemOpen
  \bibfield  {author} {\bibinfo {author} {\bibfnamefont {R.}~\bibnamefont
  {Loudon}},\ }\href@noop {} {\emph {\bibinfo {title} {The Quantum Theory of
  Light}}}\ (\bibinfo  {publisher} {OUP Oxford},\ \bibinfo {year} {2000})\
  Chap.~\bibinfo {chapter} {4}\BibitemShut {NoStop}%
\bibitem [{\citenamefont {Srednicki}(2007)}]{srednicki}%
  \BibitemOpen
  \bibfield  {author} {\bibinfo {author} {\bibfnamefont {M.}~\bibnamefont
  {Srednicki}},\ }\href@noop {} {\emph {\bibinfo {title} {Quantum Field
  Theory}}}\ (\bibinfo  {publisher} {Cambridge University Press},\ \bibinfo
  {year} {2007})\ Chap.\ \bibinfo {chapter} {1 \& 8}\BibitemShut {NoStop}%
\bibitem [{\citenamefont {Rovelli}(2007)}]{rovelli}%
  \BibitemOpen
  \bibfield  {author} {\bibinfo {author} {\bibfnamefont {C.}~\bibnamefont
  {Rovelli}},\ }\href@noop {} {\emph {\bibinfo {title} {Quantum Gravity}}}\
  (\bibinfo  {publisher} {Cambridge University Press},\ \bibinfo {year}
  {2007})\ Chap.\ \bibinfo {chapter} {5.3}\BibitemShut {NoStop}%
\bibitem [{\citenamefont {Gainutdinov}(1999)}]{QTnonloc1}%
  \BibitemOpen
  \bibfield  {author} {\bibinfo {author} {\bibfnamefont {R.~K.}\ \bibnamefont
  {Gainutdinov}},\ }\href {\doibase
  https://doi.org/10.1088/0305-4470/32/30/311} {\bibfield  {journal} {\bibinfo
  {journal} {Journal of Physics A: Mathematical and General}\ }\textbf
  {\bibinfo {volume} {32}},\ \bibinfo {pages} {5657} (\bibinfo {year}
  {1999})}\BibitemShut {NoStop}%
\bibitem [{\citenamefont {Smilga}(2017)}]{QTnonloc2}%
  \BibitemOpen
  \bibfield  {author} {\bibinfo {author} {\bibfnamefont {A.}~\bibnamefont
  {Smilga}},\ }\href {\doibase https://doi.org/10.1142/S0217751X17300253}
  {\bibfield  {journal} {\bibinfo  {journal} {International Journal of Modern
  Physics A}\ }\textbf {\bibinfo {volume} {32}},\ \bibinfo {pages} {1730025}
  (\bibinfo {year} {2017})}\BibitemShut {NoStop}%
\bibitem [{Note1()}]{Note1}%
  \BibitemOpen
  \bibinfo {note} {This can be computed in the same manner as for a simple
  harmonic oscillator \cite {feynman,rovelli}, though some extra care should be
  taken with the indices. Note that we do not use the boundary conditions $A_f
  = A_i = 0$ and the subsequent Lehmann-Symanzik-Zimmermann reduction formalism
  \cite {srednicki} here as we do not necessarily deal with scattering
  states.}\BibitemShut {Stop}%
\bibitem [{\citenamefont {Ismail}\ and\ \citenamefont
  {Simeonov}(2015)}]{complexHermite1}%
  \BibitemOpen
  \bibfield  {author} {\bibinfo {author} {\bibfnamefont {M.}~\bibnamefont
  {Ismail}}\ and\ \bibinfo {author} {\bibfnamefont {P.}~\bibnamefont
  {Simeonov}},\ }\href {\doibase 10.1090/S0002-9939-2014-12362-8} {\bibfield
  {journal} {\bibinfo  {journal} {Proceedings of the American Mathematical
  Society}\ }\textbf {\bibinfo {volume} {143}},\ \bibinfo {pages} {1397}
  (\bibinfo {year} {2015})}\BibitemShut {NoStop}%
\bibitem [{\citenamefont {Ghanmi}(2013)}]{complexHermite2}%
  \BibitemOpen
  \bibfield  {author} {\bibinfo {author} {\bibfnamefont {A.}~\bibnamefont
  {Ghanmi}},\ }\href {\doibase 10.1080/10652469.2013.772172} {\bibfield
  {journal} {\bibinfo  {journal} {Integral Transforms and Special Functions}\
  }\textbf {\bibinfo {volume} {24}},\ \bibinfo {pages} {884} (\bibinfo {year}
  {2013})}\BibitemShut {NoStop}%
\bibitem [{\citenamefont {Cotfas}\ \emph {et~al.}(2010)\citenamefont {Cotfas},
  \citenamefont {Gazeau},\ and\ \citenamefont {G{\'o}rska}}]{complexHermite3}%
  \BibitemOpen
  \bibfield  {author} {\bibinfo {author} {\bibfnamefont {N.}~\bibnamefont
  {Cotfas}}, \bibinfo {author} {\bibfnamefont {J.~P.}\ \bibnamefont {Gazeau}},
  \ and\ \bibinfo {author} {\bibfnamefont {K.}~\bibnamefont {G{\'o}rska}},\
  }\href {http://stacks.iop.org/1751-8121/43/i=30/a=305304} {\bibfield
  {journal} {\bibinfo  {journal} {Journal of Physics A: Mathematical and
  Theoretical}\ }\textbf {\bibinfo {volume} {43}},\ \bibinfo {pages} {305304}
  (\bibinfo {year} {2010})}\BibitemShut {NoStop}%
\bibitem [{Note2()}]{Note2}%
  \BibitemOpen
  \bibinfo {note} {{Note that the total probability density for emission is
  given by \begin {align*} \protect \frac {dP}{d\protect \mathcal {V}} = \DOTSI
  \intop \ilimits@ \protect \frac {d^3k}{(2\pi )^3}\left |G^{\protect \text
  {intra}}_{11 \leftarrow 00}+G^{\protect \text {inter}}_{11 \leftarrow
  00}\right |^2. \end {align*} Also, there is no need for renormalising this
  integral, as we are considering differences between the occupation in each
  state, not the total occupation number in each.}}\BibitemShut {Stop}%
\bibitem [{\citenamefont {Chen}\ \emph {et~al.}(2016)\citenamefont {Chen},
  \citenamefont {Tian}, \citenamefont {Bin-Mohsin}, \citenamefont {Nessler},
  \citenamefont {Svidzinsky},\ and\ \citenamefont {Scully}}]{paraComb3}%
  \BibitemOpen
  \bibfield  {author} {\bibinfo {author} {\bibfnamefont {G.}~\bibnamefont
  {Chen}}, \bibinfo {author} {\bibfnamefont {J.}~\bibnamefont {Tian}}, \bibinfo
  {author} {\bibfnamefont {B.}~\bibnamefont {Bin-Mohsin}}, \bibinfo {author}
  {\bibfnamefont {R.}~\bibnamefont {Nessler}}, \bibinfo {author} {\bibfnamefont
  {A.}~\bibnamefont {Svidzinsky}}, \ and\ \bibinfo {author} {\bibfnamefont
  {M.~O.}\ \bibnamefont {Scully}},\ }\href {\doibase
  10.1088/0031-8949/91/7/073004} {\bibfield  {journal} {\bibinfo  {journal}
  {Physica Scripta}\ }\textbf {\bibinfo {volume} {91}},\ \bibinfo {pages}
  {073004} (\bibinfo {year} {2016})}\BibitemShut {NoStop}%
\bibitem [{\citenamefont {Lindel}\ \emph {et~al.}(2019)\citenamefont {Lindel},
  \citenamefont {Bennett},\ and\ \citenamefont {Yoshi~Buhmann}}]{robs}%
  \BibitemOpen
  \bibfield  {author} {\bibinfo {author} {\bibfnamefont {F.}~\bibnamefont
  {Lindel}}, \bibinfo {author} {\bibfnamefont {R.}~\bibnamefont {Bennett}}, \
  and\ \bibinfo {author} {\bibfnamefont {S.}~\bibnamefont {Yoshi~Buhmann}},\
  }\href {https://arxiv.org/abs/1905.10200} {\bibfield  {journal} {\bibinfo
  {journal} {arXiv preprint arXiv:1905.10200}\ } (\bibinfo {year}
  {2019})}\BibitemShut {NoStop}%
\bibitem [{\citenamefont {Hadfield}(2009)}]{detectors}%
  \BibitemOpen
  \bibfield  {author} {\bibinfo {author} {\bibfnamefont {R.~H.}\ \bibnamefont
  {Hadfield}},\ }\href {\doibase 10.1038/nphoton.2009.230} {\bibfield
  {journal} {\bibinfo  {journal} {Nature photonics}\ }\textbf {\bibinfo
  {volume} {3}},\ \bibinfo {pages} {696} (\bibinfo {year} {2009})}\BibitemShut
  {NoStop}%
\bibitem [{\citenamefont {Agrawal}(2012)}]{agrawal}%
  \BibitemOpen
  \bibfield  {author} {\bibinfo {author} {\bibfnamefont {G.}~\bibnamefont
  {Agrawal}},\ }\href@noop {} {\emph {\bibinfo {title} {Nonlinear Fiber
  Optics}}},\ \bibinfo {edition} {5th}\ ed.\ (\bibinfo  {publisher} {Elsevier
  Inc. Academic Press},\ \bibinfo {year} {2012})\BibitemShut {NoStop}%
\end{thebibliography}

\end{document}